\documentclass[fleqn,usenatbib]{mnras}

\usepackage{newtxtext,newtxmath}

\usepackage[T1]{fontenc}

\DeclareRobustCommand{\VAN}[3]{#2}
\let\VANthebibliography\thebibliography
\def\thebibliography{\DeclareRobustCommand{\VAN}[3]{##3}\VANthebibliography}

\usepackage{graphicx}	
\usepackage{amsmath}	
\usepackage{bm} 
\usepackage{ulem} 

\newcommand{\Eq}[1]{Eq.~(\ref{eq:#1})}

\newcommand{\se}[1]{Section \ref{sec:#1}}
\newcommand{\app}[1]{Appendix \ref{app:#1}}
\newcommand{\Fig}[1]{Fig.~\ref{fig:#1}}

\newcommand{\be}{\begin{equation}}
\newcommand{\ee}{\end{equation}}
\newcommand{\bad}{\begin{equation} \begin{aligned}}
\newcommand{\ead}{\end{aligned} \end{equation}}
\newcommand{\Msun}{M_\odot}

\newcommand{\kpc}{\,{\rm kpc}}

\newcommand{\Gyr}{\,{\rm Gyr}}
\newcommand{\cm}{\,{\rm cm}}
\newcommand{\g}{\,{\rm g}}
\newcommand{\kms}{\,{\rm km/s}}

\newcommand{\rhodm}{\rho_{\rm dm}}
\newcommand{\rhodmc}{\rho_{\rm dm0}}
\newcommand{\rhob}{\rho_{\rm b}}

\newcommand{\rhos}{\rho_{\rm s}}

\newcommand{\tmerge}{t_{\text{merge}}}

\newcommand{\Mv}{M_{\rm vir}}
\newcommand{\Ms}{M_{\star}}

\newcommand{\Mb}{M_{\rm b}}

\newcommand{\MHI}{M_{\rm HI}}

\newcommand{\fb}{f_{\rm b}}

\newcommand{\rs}{r_{\rm s}}

\newcommand{\Vc}{V_{\rm c}}
\newcommand{\Vdm}{V_{\rm dm}}
\newcommand{\Vb}{V_{\rm b}}

\newcommand{\Vvir}{V_{\rm vir}}
\newcommand{\Vv}{V_{\rm vir}}

\newcommand{\sigmam}{\sigma_m}
\newcommand{\sigmaeff}{\sigma_{\rm eff}}

\newcommand{\Vmax}{V_{\rm max}}

\newcommand{\tage}{t_{\rm age}}

\newcommand{\HI}{\mathrm{H\,\textsc{i}}}
\newcommand{\pmerge}{p_{\mathrm{merge}}}
\newcommand{\hattmerge}{\hat{t}_{\mathrm{merge}}}

\title[SIDM Cross Section]{An Enhanced Isothermal Jeans Approach to Constraining Dark Matter Self-Interactions from Galactic Kinematics}

\author[Zixiang Jia et al.]{
Zixiang Jia$^{1,2}$,
Fangzhou Jiang$^{2}$\thanks{Corresponding author: fangzhou.jiang@pku.edu.cn},
Shubo Li$^{3,4,5}$,
Ran Li$^{3,5}$,
Jing Wang$^{2}$,
and
Ling Zhu$^{6}$
\\
$^{1}$Department of Astronomy, Peking University, Beijing 100871, China\\
$^{2}$Kavli Institute for Astronomy and Astrophysics, Peking University, Beijing 100871, China\\
$^{3}$School of Physics and Astronomy, Beijing Normal University, Beijing 100875, China\\
$^{4}$
National Astronomical Observatories, Chinese Academy of Sciences, 20A Datun Road, Chaoyang District, Beijing 100101, China\\
$^{5}$School of Astronomy and Space Science, University of Chinese Academy of Sciences, Beijing 100049, China\\
$^{6}$Shanghai Astronomical observatory, Chinese Academy of Sciences, Nandan road 80, Shanghai, 200030, China\\
}

\date{Accepted XXX. Received YYY; in original form ZZZ}

\pubyear{2024}

\begin{document}
\label{firstpage}
\pagerange{\pageref{firstpage}--\pageref{lastpage}}
\maketitle

\begin{abstract}
We present an improved semi-analytical model to predict density profiles of self-interacting dark matter (SIDM) halos and apply it to constrain the self-scattering cross section using SPARC galaxy rotation curves. 
Building on the isothermal Jeans approach, our model incorporates (i) velocity-dependent cross sections, (ii) an empirical treatment of core collapse, and (iii) enhanced robustness for identifying solutions. 
These advances allow us to fit a large sample of galaxies, including systems with baryon-dominated centers often excluded in earlier studies. 
We find that roughly 1/6 of galaxies admit both a core-growth and a core-collapse solution, while the rest favor a unique evolutionary state. 
Joint constraints across the sample reveal clear velocity dependence: the allowed parameter space forms an L-shaped degeneracy, where both nearly constant, low cross sections ($\sigma_0\sim2\,$cm$^2$/g,  $\omega\gtrsim500\,$km/s) and strongly velocity-dependent models ($\sigma_0\sim100\,$cm$^2$/g, $\omega\sim60\,$km/s) are viable. 
Adopting the core-growth interpretation yields best-fit values $\sigma_0\simeq5\,$cm$^2$/g and $\omega\simeq250\,$km/s. Our constraints are remarkably consistent with previous results derived from a variety of independent probes. Compared to cold dark matter (CDM) models, SIDM outperforms simple adiabatic-contraction profiles and rivals empirical feedback-based CDM profiles, yet shows no correlation with stellar-to-halo mass ratio, a proxy for feedback strength, offering a distinct explanation for dwarf galaxy diversity. Moreover, SIDM does not affect galaxy-halo scaling relations significantly and makes concentration systematically lower. Our results highlight SIDM as a compelling framework for small-scale structure, while future low-mass kinematic data will be crucial for breaking degeneracies in velocity-dependent cross-section models.
\end{abstract}




\section{Introduction}

The standard $\Lambda$CDM cosmological paradigm has been remarkably successful on cosmological large scales but continues to face challenges on subgalactic scales, particularly in explaining the structural diversity of dwarf galaxies \citep[e.g.,][]{Weinberg15, Bullock17}. 
Baryonic processes can account for some of these discrepancies \citep[e.g.,][]{Sales22}, yet self-interacting dark matter (SIDM) offers a compelling alternative that naturally alleviates small-scale tensions without requiring finely tuned baryonic feedback \citep[see][for a review]{TulinYu18}. 
In the SIDM framework, dark matter (DM) particles undergo frequent self-scattering in dense environments, driving a two-stage evolutionary process for DM halos. 
Initially, scattering thermalizes the halo center, creating a constant-density core. 
Over time, however, the isothermal core collapses as heat flows outward and the central temperature rises—an outcome of the negative heat capacity of self-gravitating systems. 
This gravothermal evolution leads to cusp–core diversity depending on halo formation times and evolutionary stage \citep[e.g.,][]{RenYu19,YangNadler23,NadlerYangYu23}. 

The pace of this evolution is governed by the DM self-interaction cross section, making it a crucial parameter for testing the SIDM paradigm. 
Observational constraints are primarily derived from galactic kinematics, especially rotation curves \citep[e.g.][]{Kaplinghat16,Kamada17,Sagunski21,Gilman21,Adhikari24,RobertsKaplinghat24,Kong25}. 
By fitting theoretical mass models to rotation curves, one can simultaneously infer halo mass, structural parameters, and the degenerate combination of halo age and cross section. 
Such models often rely on computationally expensive full $N$-body SIDM simulations \citep[e.g.][]{Rocha13,Elbert15,Robles17} or 1D gravothermal fluid simulations \citep[e.g.][]{Balberg2002, Pollack15, Essig19, Nishikawa20}. 
These methods are usually computationally costly, or relying on simplifying assumptions such as neglecting baryonic effects. 

A more efficient alternative is the isothermal Jeans model  \citep{Kaplinghat14,Kaplinghat16}.
This semi-analytic method modifies a baseline CDM halo by solving the isothermal Jeans–Poisson equation for the inner core and smoothly connecting it to the CDM-like outer halo where scattering is negligible. 
Despite its simplicity, the model reproduces simulation results with surprising accuracy \citep{Robertson20}.
It can also incorporate baryonic effects by including a baryonic potential \citep{Kaplinghat16} and modeling adiabatic contraction \citep{Jiang23}. 
However, the method struggles to capture the second phase of SIDM evolution -- gravothermal core-collapse -- because the definition of the isothermal radius, valid during core growth, breaks down once collapse begins.

Recent work has proposed an empirical workaround. 
\cite{YangJiang24} showed that the high-density solution to the isothermal Jeans equations, typically discarded, can approximate the core-collapse stage when mirrored around the point where low- and high-density solutions merge.
This approach is computationally lightweight and has been successfully applied to compact systems such as the strong lens SDSSJ0946+1006 \citep{Li25}.

In this study, we build upon the improved implementations of \citet{Jiang23} and \cite{YangJiang24}. 
Specifically, we 
(i) incorporate velocity-dependent cross sections following \citet{Yangyu22},
(ii) enhance the numerical robustness needed to identify multiple solutions and capture core collapse, and
(iii) embed these improvements in a forward-modeling framework to fit SPARC \citep[Spitzer Photometry \& Accurate Rotation Curves;][]{Lelli16} galaxies.
These advances allow us to analyze a substantially larger sample of galaxies, including systems with baryon-dominated centers that were previously excluded.

This paper is organized as follows. 
\se{model} describes our improvements to the isothermal Jeans model. 
\se{fitting} details our fitting methodology. 
\se{MassModel} presents the inferred constraints on the SIDM cross section. 
\se{discussion} examines the scatter in our constraints, compares our results with previous studies, assesses the merits of introducing SIDM relative to CDM models that include baryonic effects, and explores the impact of SIDM mass models on galaxy–halo scaling relations.
\se{conclusion} summarizes our findings.  
Throughout, we adopt a flat cosmology with  $h=0.71,\, \Omega_{\text{m}}=0.266,\, \Omega_{\Lambda}=0.734,\, \Omega_{\text{b}}=0.0465,\, \sigma_8=0.801,\,$and $ n_s=0.963$ \citep{Wmap7}.

\section{Upgraded Isothermal Jeans model}
\label{sec:model}

In this section, we present improvements to the isothermal Jeans model, building on the framework of \citet{Jiang23}. After briefly reviewing the standard formulation (\se{isomodel}), we introduce modifications to incorporate velocity-dependent cross sections (\se{vel-dependent}) and describe our empirical treatment of gravothermal core-collapse (\se{core-collapse}).

\subsection{The isothermal Jeans model}
\label{sec:isomodel}

Following \citet{Kaplinghat14,Kaplinghat16}, the SIDM effect is modeled as a modification to a CDM halo, by considering two regimes: an inner part that becomes isothermal because of frequent self-scattering, and an outer part where self-interactions are negligible and the halo remains CDM-like. 
The transition occurs at a characteristic radius, $r_1$, defined such that a typical DM particle undergoes one scattering event over the halo age, $\tage$:
\begin{equation}
    \label{eq:r1}
    \frac{4}{\sqrt{\pi}}\rho_{\text{dm}}(r_1)v(r_1)\sigmam=\frac{1}{t_{\text{age}}},
\end{equation}
where $\rhodm(r)$ is the DM density profile, $v(r)$ is the 1D velocity dispersion of the halo, and $\sigmam$ is the self-interaction cross section per unit mass. 

Inside $r_1$, the halo is assumed to be isothermal and governed by the spherical Jeans equation,
\begin{equation}
    \label{eq:Jeans}
    \frac{\mathrm{d}(\rho_{\mathrm{dm}}v^{2})}{\mathrm{d}r}+\frac{2\beta}{r}\rho_{\mathrm{dm}}v^{2}=-\rho_{\mathrm{dm}}\frac{\mathrm{d}\Phi}{\mathrm{d}r},
\end{equation}
where $\Phi$ is the gravitational potential, $\beta$ is the velocity anisotropy assumed to be zero, and $v(r)=v_0$ is the constant velocity dispersion in the isothermal region. 
The resulting density profile is
\begin{equation}
    \label{eq:SolOfPoisson}
    \rho_{\text{dm}}(r)=\rho_{\text{dm}0}\,\mathrm{exp}\left[ -\frac{\Phi(r)-\Phi(0)}{v_0^2}\right],
\end{equation}
where $\rho_{\text{dm0}}$ is the central DM density. 
The inner halo profile is then solved together with the Poisson equation,
\begin{equation}
    \label{eq:Poisson}
    \frac{1}{r^2}\frac{\mathrm{d}}{\mathrm{d}r}(r^2\frac{\mathrm{d}\Phi}{\mathrm{d}r})=4\pi G(\rho_{\text{dm}}+\rho_{\text{b}}),
\end{equation}
where $\rhob$ is the baryon density distribution, which is regarded as an input quantity inferred from observations. 

There are two free parameters in the Jeans-Poisson equations: the central density, $\rhodmc$, and the central velocity dispersion, $v_0$. 
Each pair uniquely specifies a density profile.
These parameters are constrained by treating the outer halo as a boundary condition: beyond $r_1$, where scattering is rare, the density profile must asymptotically match that of the target CDM halo.
Accordingly, $\rhodmc$ and $v_0$ are determined by minimizing the mismatch between the isothermal solution and the CDM profile at $r_1$. 
In practice, this is achieved by minimizing the objective function,
\begin{equation}
    \delta^2= \left[\frac{\rho_{\text{iso}}(r_1)-\rho_{\text{cdm}}(r_1)}{\rho_{\text{cdm}}(r_1)}\right]^2+\left[\frac{M_{\rm iso}(r_1)-M_{\text{cdm}}(r_1)}{M_{\text{cdm}}(r_1)}\right]^2,
\end{equation}
where $M(r)$ is the enclosed mass, and the subscripts `iso' and `cdm' refer to the SIDM isothermal solution and the original CDM profile, respectively. 

We refer to this procedure as {\it stitching} an isothermal SIDM core onto the CDM outskirts. 
The stitching typically yields two solutions, corresponding to distinct local minima of $\delta^2$.
These solutions share similar $v_0$ but differ in central density  $\rho_{\text{dm0}}$, and are therefore classified as the low-density and high-density branches. 
Among the two, the low-density solution is generally preferred, as it more closely matches the original CDM profile.

\citet{Jiang23} improved the workflow by incorporating adiabatic contraction of the target CDM halo. 
Specifically, they first compute the contracted CDM density profile in response to the baryonic distribution, $\rhob$, following the method of \citet{Gnedin04}.
The same stitching procedure is then applied to this contracted halo.
As in \citet{Jiang23}, we approximate the baryonic distribution with a Hernquist profile,
\begin{equation}
    \label{eq:Hernquist}
    \rho_{\text{b}}(r)=\frac{M_{\text{b}}/2\pi a^3}{(r/a)(1+r/a)^3},
\end{equation}
where $M_{\text{b}}$ is the total baryon mass and $a$ is the scale radius. 
Unlike \citet{Jiang23}, to expedite the adiabatic contraction calculation, we first compute $r_1$ using \Eq{r1} and the initial CDM halo, and then apply the \citeauthor{Gnedin04} calculation to the mass shell enclosed by $r_1$ to obtain the contracted $r_1$ radius.

\subsection{Velocity-dependent cross section}
\label{sec:vel-dependent} 

\begin{figure}
    \centering
    \includegraphics[width=0.8\linewidth]{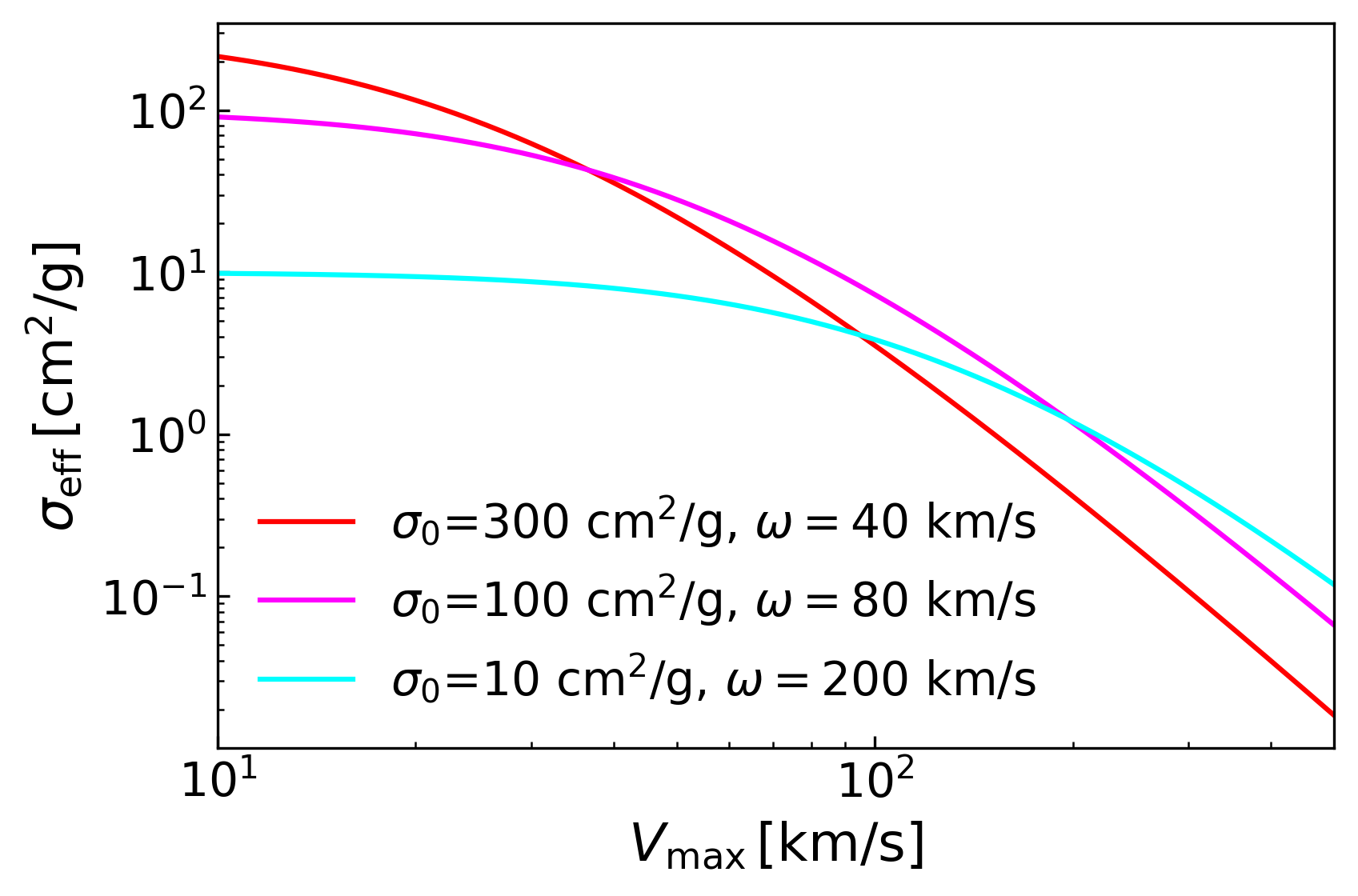}
    \caption{
    Illustration of effective cross section $\sigmaeff$ as a function of the maximum circular velocity $\Vmax$ of a halo, for velocity-dependent cross section models with different $\sigma_0$ and $\omega$, as defined in \Eq{diff-cross-section}.
    }
    \label{fig:CrossSectionModel}
\end{figure}

Previous iterations of the isothermal Jeans method typically assumed constant cross section. 
To incorporate the well-established velocity dependence, we consider elastic Rutherford scattering between dark matter particles interacting through a Yukawa potential. 
The corresponding differential cross section per unit mass in the center-of-momentum frame is
\begin{equation}
    \label{eq:diff-cross-section}
    \frac{\mathrm{d}\sigma_{m}}{\mathrm{d}\mathrm{cos}\theta}=\frac{\sigma_{0,m}\omega^4}{2[\omega^2+v^2\mathrm{sin}^2(\theta/2)]^2}.
\end{equation}

Following \citet{Yangyu22}, we encapsulate this velocity dependence in an effective cross section $\sigma_{\mathrm{eff},m}$, denoted as $\sigmaeff$ below,
\begin{equation}
    \label{eq:EffectiveCrossSection}
    \sigma_{\mathrm{eff}}=\frac{1}{2}\int\tilde{v}^{2}\mathrm{d}\tilde{v}\sin^{2}\theta\mathrm{d}\cos\theta\frac{\mathrm{d}\sigma_{m}}{\mathrm{d}\cos\theta}\tilde{v}^{5}e^{-\tilde{v}^{2}},
\end{equation}
where $\tilde{v}\equiv v/(2v_{\mathrm{eff}})$, and $v_\text{eff}$ is the characteristic one-dimensional velocity dispersion.
For an isotropic velocity field, $v_\text{eff}\approx V_{\text{max}}/{\sqrt{3}}$, where $V_{\text{max}}$ is the maximum circular velocity of the input NFW halo. 
The inner velocity dispersion is affected by baryons and can deviate from the characteristic value \citep{ZhongYang23}. In Appendix \ref{apdx: baryonic effects on the  velocity distribtion}, we show that this effect is secondary in our work.
The cross-section model is set by two parameters, $\sigma_{0,m}$, denoted as $\sigma_0$ below, the low-velocity normalization per unit mass, and $\omega$, the velocity scale above which the cross section falls sharply. 
\Fig{CrossSectionModel} illustrates the resulting dependence of $\sigmaeff$ on $\Vmax$ for various ($\sigma_0$, $\omega$).
In practice, given a choice of ($\sigma_0$, $\omega$), and a target CDM halo profile, we compute the corresponding $\sigmaeff$ and substitute it for the constant $\sigmam$ in the original workflow described in \se{isomodel}.

\subsection{Gravothermal core-collapse}
\label{sec:core-collapse} 

\begin{figure*}
    \centering
    \includegraphics[width=\textwidth]{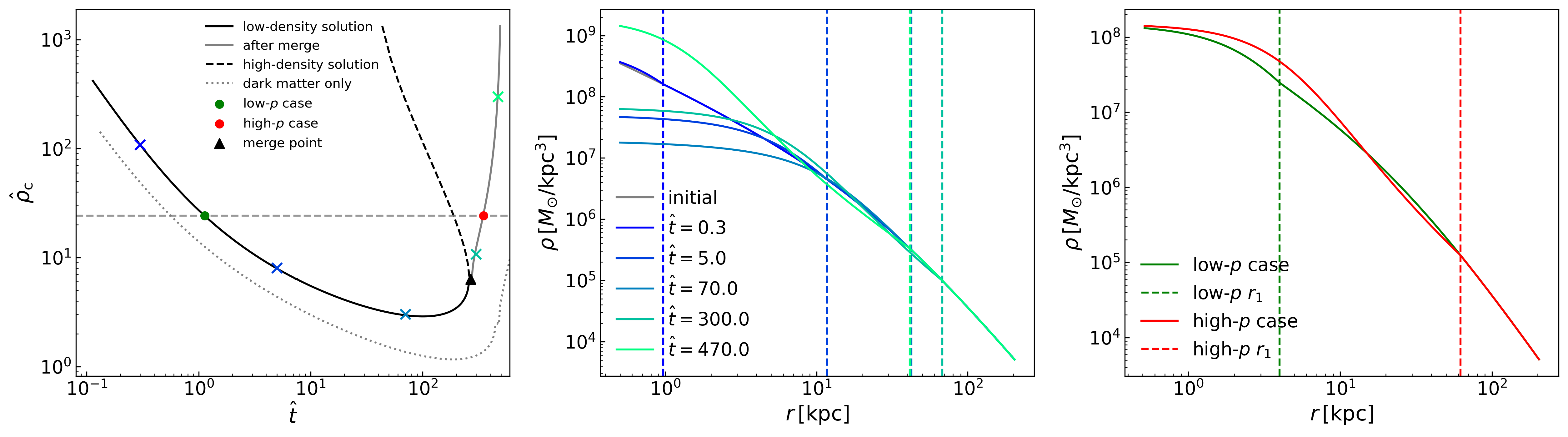}
    \caption{
    Gravothermal evolution of an SIDM halo modeled with the improved isothermal Jeans method, including an empirical treatment of core-collapse. 
    The example halo has a virial mass of $\Mv=10^{12}\Msun$, concentration of $c=10$, and a baryonic component described by a Hernquist profile with a total mass of $\Mb=10^{10}\Msun$ and a scale radius of $a=5\kpc$. 
    \quad {\it Left}: 
    Evolution of the central density from the low-density (solid black) and high-density (dashed black) solutions. 
    Density and time are shown in dimensionless form $\hat{\rho}_{\rm c}=\rho_{\rm c}/\rho_{\rm s}$, and $\hat{t}=8\sqrt{G\rhos}\rhos \rs \sigmaeff \tage$, where $\rhos$ and $\rs$ are the scale density and radius of the target NFW halo. 
    The black triangle marks the merger of the two solutions, indicating the onset of core-collapse. 
    The mirrored high-density branch (grey curve) provides an empirical approximation to the post-collapse evolution, as shown by  \protect\cite{YangJiang24}.
    Together, the solid curves trace the full gravothermal history.
    And compared to the dark-matter-only case (gray dashed line), the system with baryon experiences accelerated gravothermal evolution.
    \quad {\it Middle}:
    Density profiles along  the evolutionary track, marked by crosses in the left panel. 
    As $\hat{t}$ increases, the central density initially decreases as the core expands, reaching its maximum size at $\hat{t}\sim100$, before shrinking as the central density rises.
    \quad {\it Right}:
    Comparison of core-expanding and core-collapsing halos with identical central density.
    For example, at $p\equiv\sigmaeff\tage=1 \,\text{cm}^2/\text{g}\,\text{Gyr}$ and $308.5 \,\text{cm}^2/\text{g}\,\text{Gyr}$, the central densities conincide and the profiles are similar. This degeneracy underlies the bimodal posteriors found in MCMC fits of SIDM mass models to galaxy rotation curves (see \protect\se{MassModel}).
    }
    \label{fig:model}
\end{figure*}

We incorporate an empirical treatment of core-collapse into the isothermal Jeans model by exploiting the behavior of its two solutions. 
For a fixed cross section, as the halo age $\tage$ increases, the central densities $\rhodmc$ of the low- and high-density solutions gradually converge and eventually merge at a critical time, $\tmerge$ (see \Fig{model}, left). 
This marks the onset of gravothermal core-collapse, characterized by a subsequent rise in central density. 
Beyond this point, the standard isothermal Jeans model formally ceases to apply, since the region within $r_1$ is no longer strictly isothermal. 

As a practical workaround, \citet{YangJiang24} proposed approximating the post-collapse density profile at $\tage>\tmerge$ using the mirrored high-density solution at $2\tmerge-\tage$. Specifically, they generalize \Eq{r1} into: 
\begin{equation}
    \frac{4}{\sqrt{\pi}}\rho_{\text{dm}}(r_1)v(r_1)\sigma_{m}=\max\left[\frac{1}{\tage},\frac{1}{2t_{\mathrm{merge}}-\tage}\right].
\end{equation} 
They demonstrated that this mirrored solution reproduces the core-collapse evolution obtained from gravothermal fluid simulations \citep[e.g.,][]{Nishikawa20}, thus providing a computationally efficient empirical extension of the Jeans model into the collapse regime.

To employ the mirrored high-density solution for describing core-collapse, two requirements must be met: (i) the high-density solution must be identified robustly, and (ii) the merger time, $\tmerge$, must be determined accurately. 
Both tasks are technically challenging because, although the two solutions usually differ by orders of magnitude in central density, they converge rapidly as $\tmerge$ approaches (see \Fig{model}, left). 
As a result, near $\tage\sim\tmerge$, the search for one solution is inevitably contaminated by the presence of the other.
To overcome this difficulty, we refine the stitching procedure. 
We find that adopting a stringent convergence threshold of $\delta^2=10^{-5}$ ensures reliable identification of physically valid  solutions. 
Based on this criterion, we further developed a method to diagnose when a halo has entered the core-collapse phase, along with an efficient algorithm for computing $\tmerge$.

In addition, we improve the search algorithm by partitioning the search intervals and exploring the $\rhodmc-v_0$ space with multiple trials from different initial positions, which improves the recovery of both low- and high-density solutions.   
Details of these technical improvements are provided in \app{model}. 
With these upgrades, the isothermal Jeans model can now be applied across all stages of gravothermal evolution. And by considering  adiabatic contraction, the model naturally accounts for the effect of baryons on the core-collapse timescale.
\Fig{model} illustrates the evolution of an SIDM halo with our improved method. 
We note that halos in the core-growth and core-collapse phases can exhibit nearly identical density profiles when they share the same central density.
This degeneracy gives rise to bimodal posteriors encountered in MCMC fits to galaxy rotation curves, as discussed in \se{bimodality}.

\section{Fitting galaxy rotation curves}
\label{sec:fitting}

In this section, we describe our procedure for fitting galaxy rotation curves with mass models based on SIDM halos. 
\se{sample} introduces the observational sample and selection criteria,
\se{BaryonModel} details the modeling of baryonic distributions, and \se{MCMC} outlines the fitting strategy.

\subsection{Sample selection}
\label{sec:sample}

We use galaxy rotation curves (RCs) from the public SPARC database \citep{Lelli16}, which contains 175 late-type galaxies with Spitzer 3.6$\,\mu$m photometry and high-quality RCs compiled from previous  H$\mathrm{\,\textsc{i}}$/H$\alpha$ observations. 
The total circular velocity, $\Vc$, is derived from gaseous tracers.
Its baryonic contribution consists of a gas component, $V_{\text{gas}}$, obtained from the $\HI$ intensity map (with a correction for helium), and a stellar component, $V_\star$, derived from the surface photometry with an assumed mass-to-light ratio.
While \citet{Lelli16} adopt a uniform mass-to-light ratio of unity for all the galaxies, we rescale $V_\star$ using a more realistic estimate, as discussed in \se{BaryonModel}.  
 
To ensure robust halo mass constraints, we discard galaxies with incomplete RC coverage (fewer than eight data points) or without a clearly defined asymptotically flat region, following \citet{Lelli16}.  
For some galaxies, their rotation curves marginally pass the selection criteria of \citet{Lelli16} of showing asymptotically flat region, but the actual data quality is still not good enough for constraining their total masses -- manifested by the best-fit virial velocity $\Vvir$ being higher than the maximum observed rotational velocity $\Vmax$.
We exclude such cases with $\Vv/\Vmax\geq1$, where $\Vv$ is the virial velocity from the MCMC fitting detailed in \se{MCMC}.
We model the baryonic distribution with a single Hernquist profile, and therefore exclude galaxies showing clear evidence of two stellar components (disk and bulge). 
We further remove systems whose baryonic profiles are poorly fit by a Hernquist model.
We end up with 68 galaxies -- still among the largest kinematic datasets available for constraining SIDM models.

\subsection{Modeling baryon distributions}
\label{sec:BaryonModel}

The efficiency of the isothermal Jeans model relies in part on having an analytical form for the baryonic distribution. 
We adopt the Hernquist profile as our default description. 
This choice is motivated both by its wide applicability to dwarf and disc galaxies, and by our use of the \citet{Gnedin04} method for halo contraction, which assumes a Hernquist baryon distribution for computational efficiency.

The baryonic component includes both gas and stars.
The gas mass distribution is directly obtained from $\HI$ observations and is readily available in the SPARC database. 
For the stellar component, the observed light profile must be converted into a stellar mass profile via the mass-to-light ratio, $\Gamma$. 
However, $\Gamma$ is uncertain and may vary significantly from one galaxy to another. Since our primary interest lies in dark matter parameters, we determine $\Gamma$ independently before fitting the rotation curves with SIDM models.
We treat $\Gamma$ as an intrinsic property of the stellar population, independent of the dark matter profile, provided the latter has sufficient flexibility.
Thus, by fitting rotation curves with a range of dark matter profiles while allowing $\Gamma$ to vary freely, the average of the best-fit values offers a reasonable approximation to the intrinsic mass-to-light ratio.

Conveniently, such fits have already been performed: \citet{LiPengfei20} analyzed each SPARC galaxy using multiple dark matter profiles -- 
including NFW \citep{Navvaro97}, coreNFW \citep{Read16}, DC14 \citep{DC14}, Lucky13 \citep{LiPengfei20}, and Einasto \citep{Einasto1965} -- with $\Gamma$ treated as a free parameter. 
We adopt their results to compute a weighted average mass-to-light ratio,
\begin{equation}
    \Gamma = \frac{\sum \Gamma_i/\sigma_i^2 }{\sum 1/\sigma_i^2},
\end{equation}
where the subscript $i$ represents the fitting results from different profiles, and the summation indicates that these are summed together.
$\Gamma_i$ is the best-fit value for each case, and $\sigma_i$ is the associated uncertainty. 
In Appendix \ref{apdx: mass-to-light ratio}, we show that our mass-to-light ratio is generally consistent with the mean value suggested by stellar population synthesis  models.

Finally, we fit a Hernquist circular-velocity profile to the total baryonic rotation curve,
\begin{equation}\label{eq:Vb}
    V_{\text{b}}^2=V_{\text{gas}}^2+ \,V_\star^2,
\end{equation}
We estimate the observed total baryon mass as $M_{\mathrm{b,obs}}=\Ms+M_{\mathrm{gas}}$, where the total stellar mass is $\Ms=\Gamma L_{[3.6]}$, with $L_{[3.6]}$ the total luminosity at 3.6 $\mu$m, and the total gas mass is $M_{\mathrm{gas}}=1.33\MHI$, with $\MHI$ the total $\HI$ mass \citep{Lelli16}. 
We require the model baryon mass not deviating from $M_{\mathrm{b,obs}}$ by more than 0.5 dex.
We restrict our analysis to galaxies whose baryonic distributions are well described by the Hernquist model, excluding the 20\% of systems with the largest reduced $\chi_{\nu}^2$ from the Hernquist fits. 

\subsection{Fitting the SIDM profile}
\label{sec:MCMC}

Given the baryonic distribution, an SIDM halo is characterized by three free parameters: virial mass $\Mv$, concentration $c$, and the product of effective cross section and halo age, $\sigmaeff\tage$, which we denote as $p$. 
The likelihood function is defined as 
\begin{equation}
\label{eq:likelihood}
 \ln\mathcal{P} (\bm{\theta}| \Mv,c,p)
 =-\frac{1}{2}\sum_{i=0}^{N}\left[\frac{V_{\mathrm{c},i}-V_{\mathrm{c}}(R_{i}| \Mv,c,p)}{\sigma_{V_{\mathrm{c},i}}}\right]^{2},
\end{equation}
where $\bm{\theta}$ represents the observational data: the measured circular velocity $V_{\mathrm{c},i}$ and its uncertainty $\sigma_{V_{\mathrm{c},i}}$ at radius $R_i$. 
The model circular velocity is computed as
\begin{equation}
    V_{\text{c}}^2 = V_{\text{dm}}^2+V_{\text{b}}^2, 
\end{equation}
where $\Vb$ is the observed baryonic contribution given by \Eq{Vb}, and $\Vdm$ is is the SIDM halo contribution predicted by the isothermal Jeans model. 
We set the likelihood function to zero when $r_1$ is smaller than the radius of the innermost data point in the observed rotation curve.

The posterior distribution of the model parameters,  $\mathcal{P}(\Mv,c,p|\bm{\theta})$, is given by Bayes' theorem:
\begin{equation}
\label{eq:posterior of halo parameters}
\mathcal{P}(\Mv,c,p|\bm{\theta})\propto\mathcal{P}(\bm{\theta}|\Mv,c,p)\mathcal{P}(\Mv,c,p),
\end{equation}
where $\mathcal{P}(\Mv,c,p)$ denotes the prior. 
We adopt a uniform prior for the logarithmic halo mass
\be
\log\frac{\Mv}{\Msun} \in\left[\mathrm{max}\left\{\log \frac{\Mb}{\Msun}+0.3,9\right\},\mathrm{min}\left\{\log\frac{\Mb}{\Msun}+4,13\right\}\right].
\ee
The bounds are motivated by two considerations: (i) virial masses typically exceed baryonic masses by at least 0.3 dex, consistent with the cosmic baryon fraction; and (ii) since SPARC galaxies are predominantly late-type systems, their halos are unlikely to be more massive than those of galaxy groups.
For the product of halo age and effective cross section $p$, we adopt a uniform prior on log scale, 
\be
\log (p/[\text{cm}^2/\text{g}\,\text{Gyr}]) \in [-1, 4].
\ee
The lower bound corresponds to a young halo ($\tage \sim 0.3 \Gyr$) with a small cross section (0.3$\,\cm^2/\g$), while the upper bound corresponds to an old halo with Hubble-time age and a very large cross section (1000$\,\cm^2/\g$). 
These limits safely encompass all realistic combinations.
For halo concentration, we adopt a log-normal prior motivated by the well-established concentration–mass relation, with the median set by a power law and a scatter of 0.11 dex \citep[e.g.,][]{DM14}. 

We sample the posterior distributions of $\log\Mv$, $\log c$, and $\log p$ using the Markov Chain Monte Carlo (MCMC) method implemented with the ensemble sampler in {\tt emcee} \citep{emcee}. 
We employ 50 walkers to explore the parameter space with the {\tt StretchMove} algorithm, running each chain for 3000 steps and discarding the first 600 steps as burn-in.
To mitigate rare cases where some walkers become trapped in low-posterior regions, we exclude those with anomalously small posterior values.

The determination of $\tmerge$ during model evaluations is computationally expensive. 
To accelerate the process, we pre-compute lookup tables of $\tmerge$ for each galaxy on a grid spanned by $\log(\Mv/\Msun)$ and $c$, given its baryonic distribution. 
\footnote{The grid covers $\mathrm{max}\{\log(\Mb/\Msun)+0.3,9\}<\log(\Mv/\Msun)<\mathrm{min}\{\log(\Mb/\Msun)+4,13\}$ and $4<c<40$, with 10 evenly spaced points in each dimension.} 
For each model evaluation, $\tmerge$ is then obtained by cubic-spline interpolation from these pre-computed tables.

The posterior distributions are sometimes bimodal, indicating that either a core-growth or a core-collapse solution can reproduce the data. 
This is an important aspect of our results and will be discussed further in \se{bimodality}. 
To detect and separate bimodal posteriors, we first apply the {\tt K-means} algorithm in {\tt scikit-learn} \citep{sklearn} to divide the MCMC samples into two clusters.
We then assess whether the posterior indeed contains two physically distinct peaks in terms of the parameter $p$.
Specifically, we compute the mean ($\mu_i$) and the standard deviation ($\sigma_i$) of $\log p$ within each cluster ($i=1,2$), and classify the posterior as bimodal if $|\mu_1-\mu_2|>2(\sigma_1+\sigma_2)$. 
Visual inspection confirms that this procedure reliably identifies bimodality in $p$ for all applicable galaxies.

\section{SIDM mass models of SPARC galaxies}
\label{sec:MassModel}

In this section, we present SIDM mass models for the selected SPARC galaxies, focusing on constraints on the self-interaction cross section, or more precisely, the degenerate product $p\equiv\tage\sigmaeff$. 
\se{bimodality} highlights case studies with unimodal and bimodal posteriors. 
\se{constraints} then derives effective cross sections  $\sigma_{\text{eff}}$ by making physical assumptions on halo age, and performs a joint analysis across the full sample to constrain $\sigma_0$ and $\omega$.

\begin{figure*}
    \centering
    \includegraphics[width=\textwidth]{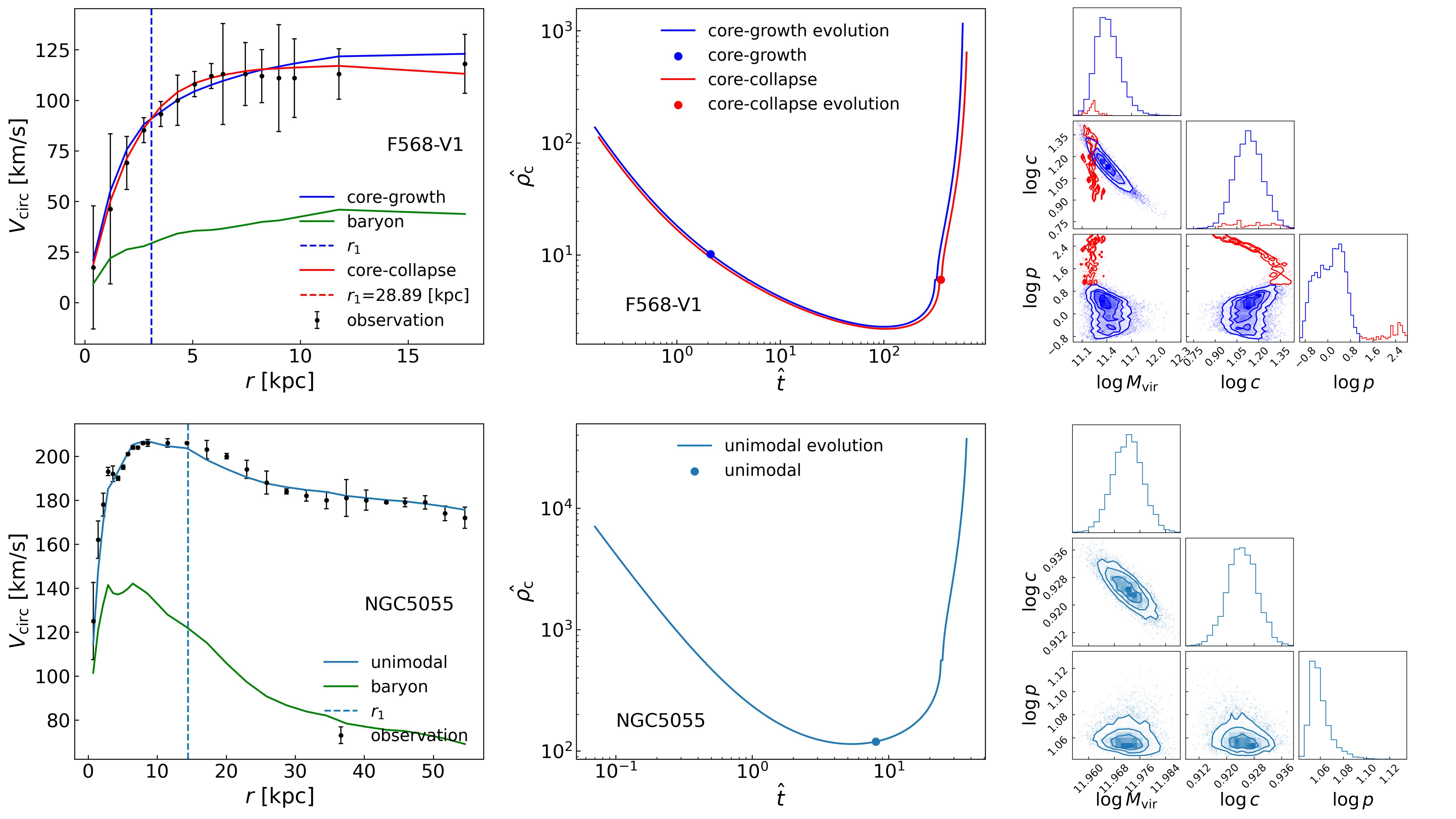}
    \caption{
    Representative case studies of SIDM rotation-curve fitting. 
    \quad
    {\it Left}: 
    Observed circular-velocity profiles (black points with 1$\sigma$ error bars) from the SPARC survey \citep{Lelli16}. 
    The baryonic contribution (green curve) is derived from $\mathrm{H\textsc{i}}$ intensity and stellar light profiles, with mass-to-light ratios determined as described in \se{BaryonModel}.
    The offset between the data points and the baryonic curve corresponds to the DM contribution. 
    The total mass model, combining the baryonic distribution with the best-fit SIDM halo, is shown by the blue (and red) curves.
    \quad
    {\it Middle}: Evolution of the central density of the best-fit SIDM halos. 
    The filled circles indicate the evolutionary stage that provides the best fit.
    \quad
    {\it Right}: Posterior distributions of the model parameters: halo virial mass $\Mv$, concentration $c$, and the parameter $p=\tage\sigmaeff$, i.e. the product of halo age and effective cross section.
    \quad
    The {\it upper} row illustrates a case with bimodal posterior: both a core-growth solution ($\Mv=10^{11.4}\Msun$, $c=13.5$, and  $p=1.31\cm^2/\g\Gyr$) and a core-collapse solution ($\Mv=10^{11.2}\Msun$, $c=14.6$, and  $p=202\cm^2/\g\Gyr$)  provide comparably good fits, differing by more than an order of magnitude in $p$.
    This bimodality reflects degeneracy in the cross section rather than halo age (see \se{bimodality}).
    The {\it lower} row shows an example with a unimodal posterior: a halo with $\Mv=10^{12.0}\Msun$, $c=8.41$, and $p=11.4\cm^2\g^{-1}\Gyr$.
    This halo lies near the maximal-core stage of gravothermal evolution, as indicated by the circle in the central-density track.
    }
    \label{fig:examples}
\end{figure*}


\begin{figure*}
    \centering
    \includegraphics[width=0.8\textwidth]{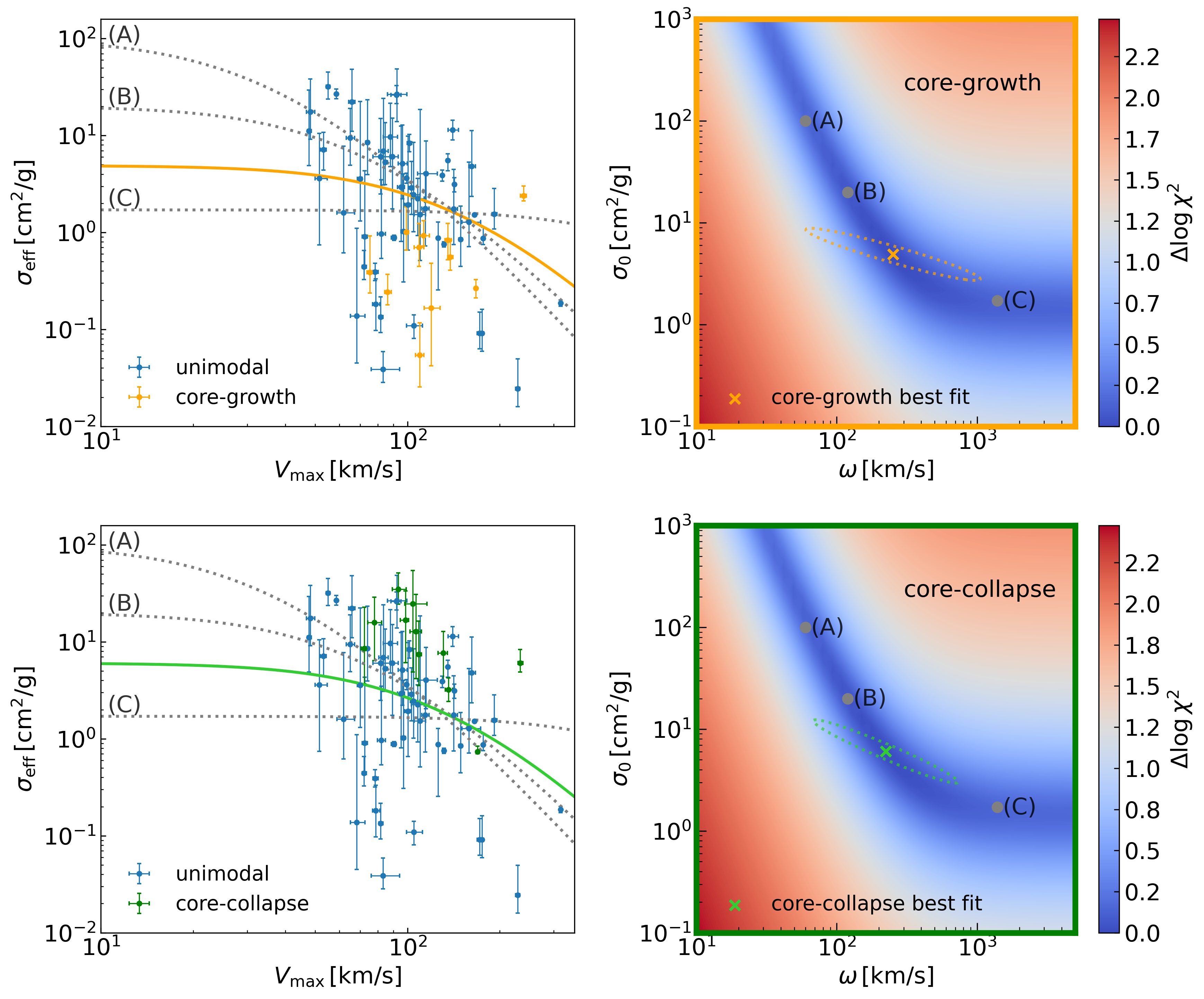}
    \caption{
    Maximum circular velocities ($\Vmax$) and effective cross sections ($\sigmaeff$) inferred from SIDM mass-model fits to galaxy rotation curves ({\it left}), and the resulting joint constraints on the cross-section parameters $\sigma_0$ and $\omega$ ({\it right}). 
    Blue symbols denote galaxies with unimodal posterior.
    Orange and green symbols represent the core-growth and core-collapse solutions, respectively, for galaxies with bimodal posteriors. Error bars indicate 68\% confidence intervals.
    \quad
    The joint constraints on $\sigma_0$ and $\omega$ are obtained by fitting the effective cross section defined in \Eq{EffectiveCrossSection} to the ($\Vmax$, $\sigmaeff$) data. 
    In the {\it right} panels, we adopt the core-growth and core-collapse solutions, respectively, for bimodal galaxies, while including unimodal galaxies in both analyses. 
    The color scale indicates the $\chi^2$ values relative to the minimum  in the $\sigma_0$-$\omega$ space (\Eq{chisquare}). 
    Constraints on the parameters do not significantly depend on the choice of solutions. 
    With either core-growth or core-collapse solutions, the allowed region forms a clear L-shaped band, where both high-$\sigma_0$ low-$\omega$ and low-$\sigma_0$ high-$\omega$ models are comparably viable. 
    Three representative examples (A, B, and C) illustrate this degeneracy.
    When adopting core-growth solutions, the best-fit MCMC model lies between (B) and (C), with $\sigma_0 = 4.9\cm^2/\g$ and $\omega=250\kms$, marked by an orange cross and a surrounding contour approximating the 3$\sigma$ posterior interval; the corresponding $\sigmaeff(\Vmax)$ relation is shown by the orange curve. 
    \quad
    When adopting core-collapse solutions instead, the parameter constraints tighten and the L-shaped allowed region narrows slightly.
    The best-fit model is similar, given by $\sigma_0 = 6.0\cm^2/\g$ and $\omega=220\kms$, marked by a green cross with a $3\sigma$ contour, and its $\sigmaeff(\Vmax)$ relation is shown by the green curve. 
    \quad
    Without kinematics data for ultra-faint dwarfs ($\Vmax\lesssim 30\kms$), these degeneracies remain difficult to break.
    }
    \label{fig:constraints}
\end{figure*}

\subsection{Existence of core-collapsing SIDM halos}
\label{sec:bimodality}

Roughly 1/6 of the sample (11 galaxies) exhibits double-peaked posteriors.  
An example is shown in the upper panels of \Fig{examples}. 
This bimodality arises because a halo in the core-growth phase can produce a density profile nearly indistinguishable from that of a core-collapsing halo at later times (see \Fig{model}). 
As a result, both solutions are consistent with the observed rotation curves.

Since $p$ is the product of halo age $\tage$ and effective cross section $\sigmaeff$, a key question for bimodal galaxies is whether the bimodality originates from variations in $\sigmaeff$ or in $\tage$. 
Clues come from comparing the separation between the $p$ values of the two posteriors with expectations from halo formation-time distributions (see \se{constraints}).

Typically, the two $p$ values differ by more than an order of magnitude, while the difference in the two $\tage$ values is typically less than an order of magnitude. For example, for a halo of mass $\Mv=10^{13}\Msun$, the mean lookback time to formation is $6.6\Gyr$, while the probability of such a halo forming within the past 1 Gyr is very small, $P(\tage<1\Gyr|\Mv)=0.0146$. 
Thus, such large differences in $p$ are unlikely to arise from  variations in halo age, and instead imply bimodality in the cross section.
For the purpose of constraining cross section, we therefore retain both solutions and explore two scenarios in the following analysis: one in which the low-$p$ solution is adopted, and one in which the high-$p$ solution is taken.

For galaxies with unimodal posteriors, we find that nearly half of them are located near the stage of maximum core size.  
An example is shown in the lower panels of \Fig{examples}. 
The rest of the unimodal sample have only core-growth solutions.

\subsection{SIDM cross section}
\label{sec:constraints}

To infer SIDM cross sections from the fitted values of $p$, we must assume a halo age, $t_{\text{age}}$, for each galaxy. 
In the absence of direct observational constraints on the assembly histories of SPARC galaxies, we estimate their average formation times as a function of halo mass using extended Press–Schechter theory \citep[e.g.,][]{LaceyCole93}. 

For a halo of mass $M_0$ at redshift $z_0$, the probability that it assembled half of its mass by redshift $z_1$ is
\begin{equation}
    \mathscr{P}(>z_\text{1}|M_0,z_0)=\int_{M_0/2}^{M_0}\frac{M_0}{M_1}f_\text{FU}(S_1,\delta_1|S_0,\delta_0)\left|\frac{\mathrm{d}S_1}{\mathrm{d}M_1}\right|\mathrm{d}M_1,
\end{equation}
where $S\equiv\sigma^2(M)$ is the variance of the linear density field on  mass scale $M$, and $\delta\equiv 1.686/D(z)$ is the critical linearized overdensity for spherical collapse with $D(z)$ the linear growth rate, and the conditional progenitor mass distribution is
\begin{equation}
    f_{\mathrm{FU}}(S_1,\delta_1|S_0,\delta_0) =\frac{1}{\sqrt{2\pi}}\frac{\delta_1-\delta_0}{(S_1-S_0)^{3/2}}\exp\left[-\frac{(\delta_1-\delta_0)^2}{2\left(S_1-S_0\right)}\right].
\end{equation}
We adopt the standard halo formation time as the epoch when the halo has assembled half of its present mass, such that $\mathscr{P}$ represents the cumulative probability distribution of formation redshifts. 
Using the implementation of this formalism in the public code {\tt SatGen}\footnote{\url{https://github.com/JiangFangzhou/SatGen}} \citep{SatGen}, we compute the mean formation redshift for halos of a given mass, and convert it into the corresponding lookback time, which we take as $\tage$. 

For each posterior sample $(\log\Mv, \log c, \log p)$, we compute $\tage$ using the method described above and obtain the effective cross section $\sigmaeff$ as $\sigmaeff=p/\tage$. 
We also calculate the maximum circular velocity $\Vmax$ of the initial NFW halo. 
By randomly sampling the posterior distributions, we derive the median values and the 68\% confidence intervals for both $\sigmaeff$ and $\Vmax$.

The results are shown in the $\sigma_{\text{eff}}-V_{\text{max}}$ plane in \Fig{constraints}, where several representative velocity-dependent cross-section models $(\sigma_0, \omega)$ are overplotted for comparison.
Overall, the galaxies display substantial scatter, with best-fit effective cross sections spanning a wide range. No single cross-section model can account for the full diversity of the sample.

We further constrain the two parameters, $\sigma_0$ and $\omega$, in the velocity-dependent cross-section model (\Eq{EffectiveCrossSection}), using the inferred values of $\Vmax$ and $\sigmaeff$ for all galaxies. 
We treat these values as data points, accounting for the uncertainties in $\sigmaeff$.\footnote{For simplicity, we neglect the uncertainties in $\Vmax$.}
Although no single cross-section model can reproduce all galaxies simultaneously, we can still ask: which models best minimize the overall disagreement? 
To quantify this, we define the objective function
\begin{equation}\label{eq:chisquare}
    \chi^2 =\sum_{i}\left[\frac{\log\sigma_{\text{eff}}(V_{\text{max},i}|\sigma_0,\omega)-\log\sigma_{\text{eff},i}}{\sigma_{\log\sigma_{\text{eff}},i}}\right]^2,
\end{equation}
where $\sigma_{\text{eff}}(V_{\text{max},i}|\sigma_0,\omega)$ is the model-predicted effective cross section for the $i$th galaxy given $\sigma_0$ and $\omega$, $\sigma_{\text{eff},i}$ is the observationally inferred value, and $\sigma_{\log\sigma_{\text{eff}},i}$ is its uncertainty.

We test whether the population is better explained if bimodal galaxies are all in the core-growth phase vs all in the core-collapse phase.
Hence, we derive the constraints on $\sigma_0$ and $\omega$ under two scenarios for handling these dual-solution systems: 
(i) adopting only the core-growth solutions, and 
(ii) adopting only the core-collapse solutions. 
In either case, these galaxies are analyzed together with those that have a single solution.

The right panels of \Fig{constraints} show the resulting objective values in the $\sigma_0$-$\omega$ plane.
More specifically, the plotted quantity is $\Delta\log\chi^2=\log\chi^2(\omega,\sigma_0)-\log\chi^2_{\text{min}}$, 
where $\chi^2_{\text{min}}$ is minimum value across the parameter space, corresponding to the best-fit model.
When adopting the core-growth solutions for bimodal systems, we find a clear anti-correlation between $\sigma_{\text{eff}}$ and $\Vmax$. 
The resulting joint constraint forms an L-shaped degeneracy in the $\sigma_0$- $\omega$ plane. 
Both strongly velocity-dependent models -- with large $\sigma_0 \sim100\cm^2/\g$ and low $\omega\sim 60\kms$ -- and nearly constant cross-section models -- with $\sigma_0\sim 2\cm^2/\g$ and $\omega\gtrsim 500\kms$ -- are comparably favored. The constraints on the parameter space do not exhibit significant dependence on the choice of solutions.
When adopting the core-collapse solutions, the constraints on the parameters are slightly strengthened, and the L-shaped allowed region becomes slightly narrower,
as shown in the lower right panel of \Fig{constraints}.

Besides scanning the parameter space to evaluate $\chi^2$, we also infer $\sigma_0$ and $\omega$ using Bayesian approach with MCMC via the {\tt emcee} ensemble sampler.
The posterior is given by
\begin{equation}
\label{eq:posterior of cross section parameters}
\mathcal{P}(\sigma_0,\omega|\bm{\Theta})\propto\mathcal{P}(\bm{\Theta}|\sigma_0,\omega)\mathcal{P}(\sigma_0,\omega),
\end{equation}
where $\bm{\Theta}$ denotes the inferred ($\Vmax$, $\sigmaeff$) values from the RC fits, $\mathcal{P}(\sigma_0,\omega)$ is the prior, and the likelihood is defined as: 
\begin{equation}
    \ln \mathcal{P}(\bm{\Theta}|\sigma_0,\omega) = -\frac{1}{2}\chi^2.
\end{equation}
We adopt uniform priors of $[-1,3]$ for $\log(\sigma_0/[\cm^2/\g])$ and $[1,3.7]$ for $\log(\omega/[\kms])$.
The posterior is sampled with 50 walkers run for 3000 steps, discarding the first 600 burn-in steps. 
The resulting posteriors in $(\log\sigma_0,\log\omega)$ are compact and well approximated by two-dimensional Gaussians, shown as $3\sigma$ contours in \Fig{constraints}.
For the core-growth cases (together with unimodal galaxies), the posterior medians are $\sigma_0=4.9\cm^2/\g$ and $\omega=250\kms$.
For the core-collapse cases (plus unimodal galaxies), the medians shift to $\sigma_0=6.0\cm^2/\g$ and $\omega=220\kms$. 
The medians do not show a significant difference, which further indicates that the choice of solutions has no substantial impact on the constraints of the parameter space.

We emphasize that no single cross-section model provides a satisfactory fit to all galaxies, and therefore the tight MCMC posteriors should not be over-interpreted. 
The L-shaped degeneracy seen in the $\chi^2$ maps reinforces this point. 
The MCMC results primarily serves the purpose of identifying the more probable regions within a degenerate parameter space.

\section{Discussion}\label{sec:discussion}

In this section, we place our results in the context of previous studies and the broader literature on the galaxy–halo connection. 

In \se{comparison}, we compare our constraints with those obtained in previous studies. 
In \se{scatter}, we investigate the origin of the large scatter in the $\sigmaeff$–$\Vmax$ plane and discuss the prospects and limitations of using individual galaxies to infer the self-interaction cross section.

A common concern regarding the SIDM paradigm is that, by introducing an additional degree of freedom relative to CDM, it may naturally reproduce the observed diversity of galaxy rotation curves. However, within the standard CDM framework, a non-negligible level of structural diversity can also arise once the response of dark-matter halos to baryonic processes is taken into account. Accordingly, in \se{SIDMvsCDM} and \se{SIDMorFeedback}, we assess the extent to which SIDM is required relative to CDM models that incorporate baryonic feedback and halo response.

Finally, in \se{ScalingRelations}, we extend the comparison to the population level by examining how SIDM modifies key scaling relations in the galaxy–halo connection. In particular, we highlight that halo masses and concentrations inferred under SIDM can differ systematically from those derived assuming CDM, with important implications for interpreting galaxy–halo correlations.

\subsection{Comparison with previous constraints}\label{sec:comparison}

\begin{figure*}
    \centering
    \includegraphics[width=0.75\linewidth]{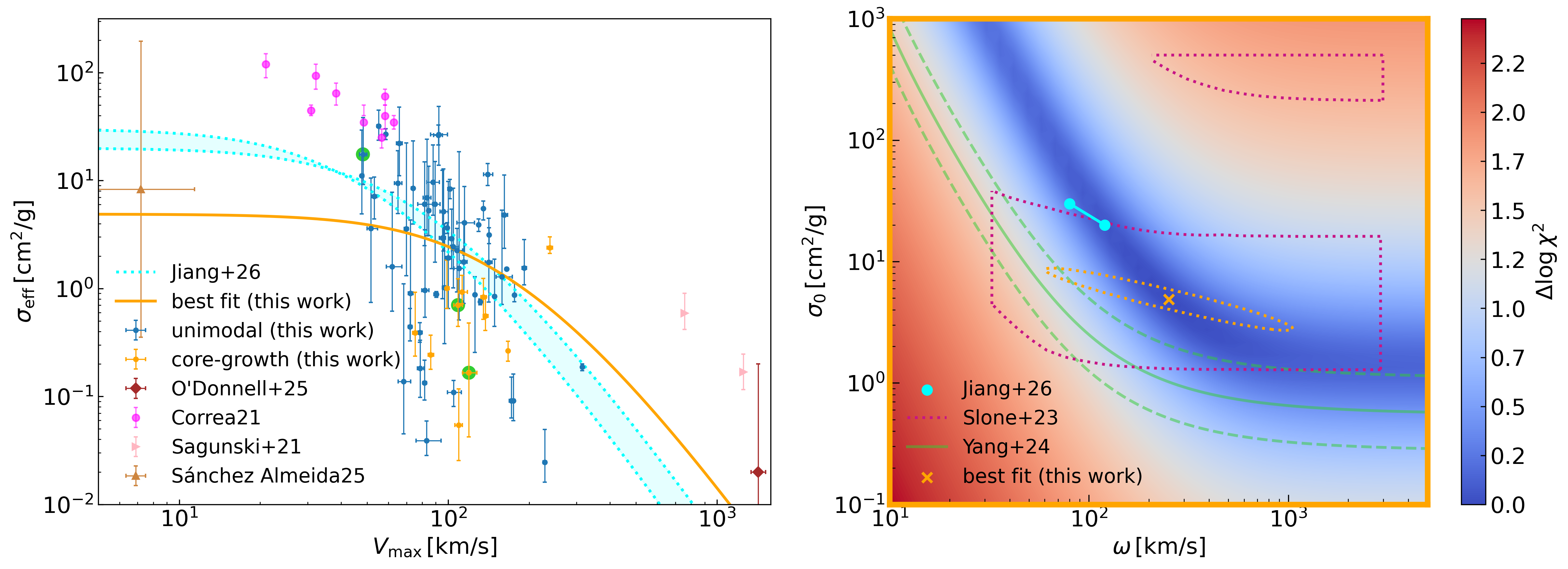}
    \caption{
    Comparison of constraints on the velocity-dependent SIDM cross section from this work with other studies across various scales.
    {\it Left}: The inferred effective cross section versus maximum circular velocity $\Vmax$ for the SPARC sample, taking the core-growth solution for bimodal cases. 
    Blue (orange) points represent unimodal galaxies (the core-growth solution of bimodal galaxies) used in this study. 
    Three of our galaxies are highlighted with green outlines, representing 1/3 of the sample analyzed by \citet[][while their remaining SPARC galaxies fail to meet our selection criteria.]{YangJiang24}.
    Points in other colors show constraints from previous studies using other probes: the stellar cores in ultra faint dwarfs \citep{SA25}, Milky Way satellites \citep{Correa21}, strong lensing and stellar kinematics of groups and clusters \citep{Sagunski21, O'Donnell25}.
    \quad
    {\it Right}: Joint constraints in the $\sigmaeff-\Vmax$ plane. 
    The blue L-shaped region is from this work. 
    Overlaid for comparison are the constraints from \citet{YangJiang24} and \citet{Slone23}. 
    The orange cross marks the best-fit parameters adopting unimodal galaxies and the core-growth solutions of bimodal galaxies, with the surrounding ellipse approximating the $3\sigma$ posterior interval, the same as in \Fig{constraints}. 
    In addition to the parameter space excluded by \citet{Slone23} (purple dotted lines), a portion of the parameter space remains ($\sigma_0\gtrsim30\cm^2/\g,\omega\lesssim100\kms$).
    Offset exists between our work and \citet{YangJiang24}, shown in green solid line with dashed lines indicating 1$\sigma$ scatter. 
    We refrain from over-interpreting this difference, as the \citeauthor{YangJiang24} sample is small and biased.
    The cyan point represents the cross sections required to reproduce the SMBH mass functions inferred for the Little Red Dots in the early Universe \citep{Jiang25}, which lies closely to our allowed region.
    }
    \label{fig:cross section comparison}
\end{figure*}

We note that the shape of the allowed region derived from our new model is consistent with previous kinematic constraints, which is based mainly on a much smaller subsample of the SPARC survey \citep{YangJiang24}.
However, a systematic offset exists between the two, as shown in the right panel of \Fig{cross section comparison}. 
We refrain from over-interpreting this difference, as the \citeauthor{YangJiang24} sample consists of only seven SPARC galaxies and two brightest cluster galaxies, which represents a biased subset of baryon-poor, spatially extended systems. 
These galaxies were selected partly because earlier versions of the isothermal Jeans model suffered from numerical robustness issues when applied to systems with compact and baryon-dominated inner regions. 
Under our updated selection criteria described in \se{sample}, we find that four of the seven SPARC galaxies do not qualify for inclusion, mainly because their rotation curves do not reach a clear plateau or maximum.
For the remaining three SPARC galaxies common to both studies, we highlight their locations in the $\sigmaeff$-$\Vmax$ plane derived from our new model, as shown in the left panel of \Fig{cross section comparison} -- two of them indeed have best-fit cross sections on the lower end, with $\sigmaeff\lesssim 1\cm^{2}/\g$.

We also compare our constraints with previous constraints based on probes other than rotation curves. 
These include limits from central density estimates based on observations of stellar cores in ultra faint dwarfs \citep{SA25} and Milky Way satellites \citep{Correa21,Slone23}, as well as constraints from strong lensing combined with stellar kinematics in groups and clusters \citep{Sagunski21,O'Donnell25}.
Compared to our results, as shown in the left panel of \Fig{cross section comparison}, these studies exhibit good overall agreement and together strongly favor a cross section that decreases monotonically with velocity.
Although part of the parameter space lies within the region excluded by \citet{Slone23} (purple dotted line), viable parameter space still remains at $\sigma_0\gtrsim30\cm^2/\g, \omega\lesssim100\kms$. 
Given the substantial scatter in the inferred effective cross section, the best-fit values of $\sigma_0$ and $\omega$ are provided for reference only. 
The fact that these reference values fall within the excluded region does not imply the SIDM model is ruled out.

More intriguingly, the allowed region in our work is remarkably close to the cross-section values required to reproduce the supermassive black hole (SMBH) mass functions inferred for the Little Red Dots in the early Universe \citep{Jiang25}, which is characterized by $\sigma_0 \sim 20-30 \cm^2/\g$ and $\omega \sim 80-120 \kms$.
In that cosmological framework, SMBHs form through the core-collapse of rare, early-forming, high-concentration SIDM halos. 
Taken together, these results suggest that the SIDM paradigm admits a region of parameter space that is simultaneously compatible with orthogonal constraints from nearby galaxy kinematics and from SMBH demographics at cosmic dawn.

\subsection{Scatter in constraints on cross section}
\label{sec:scatter}

\begin{figure*}
    \centering
    \includegraphics[width=0.85\linewidth]{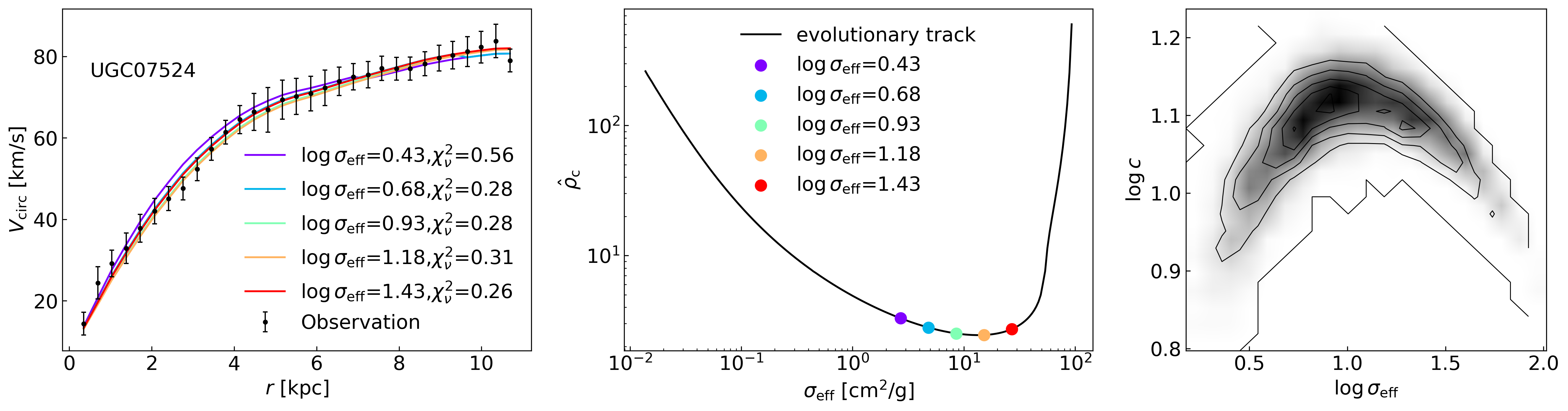}
    \caption{
    Illustration of limited constraining power on SIDM cross section for a halo near the maximal-core phase and the degeneracy between $c$ and $\sigmaeff$, using galaxy UGC07254 as an example. 
    {\it Left}: observed rotation curve (black points) of UGC07524 with best-fit SIDM model (cyan line). 
    Models with fixed halo mass, concentration, and halo age but different effective cross section values $\sigmaeff$ (colored lines) produce nearly identical rotation curves with similarly low $\chi^2_{\nu}$, demonstrating the insensitivity of individual rotation curves to the cross section during this evolutionary stage.
    {\it Middle}: gravothermal evolutionary track of the halo, in terms of normalized central density as the effective cross section, for a fixed halo age. 
    The model rotation curves of different colors on the left correspond to points of the same colors on the right, which are all near the maximal-core stage.  
    {\it Right}: the distribution of $\log c$ and $\log\sigmaeff$ from posterior sampling, where the color represents the density of samples in parameter space.
    In the core-growth phase ($\log\sigmaeff\lesssim1$), $c$ and $\sigmaeff$ are positively correlated, whereas in the core-collapse phase ($\log\sigmaeff\gtrsim1$), the correlation is reversed. This implies a significant degeneracy between $c$ and $\sigmaeff$, highlighting the challenge of precisely constraining cross section.
    }
    \label{fig:maximal core example}
\end{figure*}

Our best-fit models result in substantial scatter observed in the $\sigmaeff$–$\Vmax$ plane. 
We find that this scatter largely originates from the limited constraining power on $\sigmaeff$ from galaxies near the maximal-core stage of gravothermal evolution. 
We illustrate this point with the galaxy UGC07524, as shown in \Fig{maximal core example}. 
For this system, we keep its $\Mv$, $c$ and $\tage$ fixed at the best-fit values and vary the effective cross section $\sigmaeff$ by up to $\pm0.5$ dex. 
Within this wide range of $\sigmaeff$, the model rotation curves always remain roughly consistent with the data, because all these values correspond to a state near the maximal-core phase, during which the central density -- and hence the overall density profile -- is largely insensitive to changes in $\sigmaeff$. 
Hence, the cross section cannot be effectively constrained for {\it individual} systems near maximal core expansion. 
Approximately 40\% of galaxies in our sample are near this evolutionary stage, leading to the large scatter in $\sigmaeff$–$\Vmax$ plane.

Degeneracy with halo parameters, particularly concentration, also significantly impacts the inference of cross section. 
In Figure 2 of \citet{RobertsKaplinghat24}, they demonstrate that for a given galaxy, simultaneously varying both concentration and cross section yields rotation curves in good agreement with observations, implying a strong degeneracy between concentration and cross section. 
Such degeneracy is also commonly observed in our study. 
One example is shown in the right panel of Figure 6, which presents the distribution of $\log c$ and $\log\sigmaeff$ from posterior sampling of the galaxy UGC07254. 
At $\log\sigmaeff\lesssim1$, corresponding to the core growth phase, $c$ and $\sigmaeff$ are positively correlated; in contrast, at $\log\sigmaeff\gtrsim1$, corresponding to the core-collapse phase, the  correlation is reversed.
Notably, in \citet{RobertsKaplinghat24}, NGC3109, which resides in the core collapse regime, follows exactly the same correlation pattern as we find. 
When $c$ cannot be precisely determined, $\sigmaeff$ also cannot be robustly inferred.
This effect contributes to the large scatter in the $\sigmaeff-\Vmax$ plane.

In addition, several aspects of both the data and the modeling remain uncertain or oversimplified. 
For example, while we allow different mass-to-light ratios for different galaxies -- an improvement over earlier studies that assumed a constant value -- we do
not yet account for stellar-population gradients. 
Likewise, the modeling of baryonic distributions and adiabatic contraction can be refined with the aid of numerical simulations.
We defer such improvements to future studies. 
We conclude that  the large scatter in the $\sigmaeff-\Vmax$ distribution reflects a genuine limitation in constraining the SIDM cross section via rotation curve fitting, rather than evidence against the SIDM paradigm itself. 

\subsection{Comparison to CDM models with baryonic effects}
\label{sec:SIDMvsCDM}

\begin{figure*}
    \centering
    \includegraphics[width=1\textwidth]{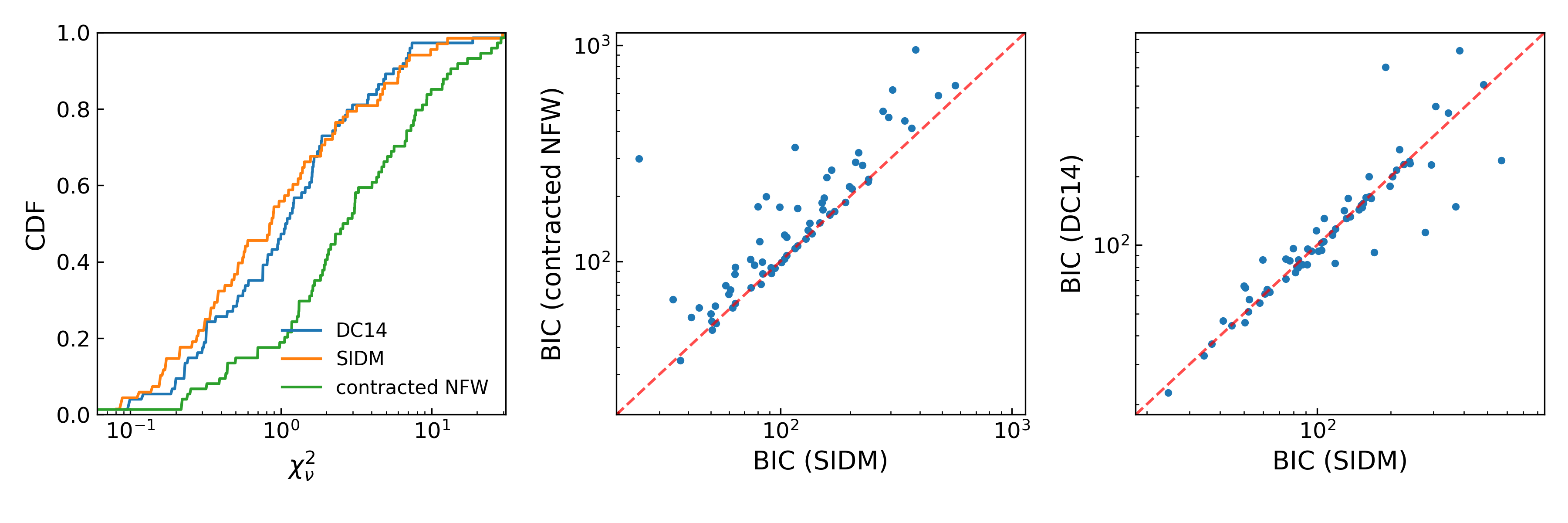}
    \caption{
    Cumulative distributions of reduced chi-square ($\chi_{\nu}^2$) ({\it left}) and Bayesian Information Criterion (BIC) comparisons ({\it middle} and {\it right}) for different models: the SIDM model developed in this study, the \citet{DC14} CDM model with an empirical treatment of baryonic effects, and a simple CDM model with  adiabatic contraction. 
    \quad
    The SIDM and DC14 models both provide significantly better fits than the adiabatic-contraction model, and have similar performance in terms of BIC. 
    }
    \label{fig:BIC}
\end{figure*}

We consider two CDM models that incorporate the structural response of halos to baryonic effects.
First, we include the effect of adiabatic contraction on an NFW halo. 
This is directly analogous to the SIDM framework, where the target NFW halo contracts in response to a baryonic distribution described by a Hernquist profile, following the prescription of  \citet{Gnedin04}. 
We refer to this as the {\it contracted NFW} model.  
Second, we adopt the empirical description of baryonic effects developed by \citet{DC14}, calibrated on hydrodynamical cosmological simulations.
This model uses the $\alpha\beta\gamma$ parameterization of the density profile, 
\begin{equation}
    \rho(r)=\frac{\rho_\mathrm{s}}{(r/\rs))^\gamma\left[1+(r/\rs)^\alpha\right]^{(\beta-\gamma)/\alpha}},
\end{equation}
where the slopes are expressed as functions of the stellar-to-halo mass ratio, $X=\log(\Ms/\Mv)$:
\begin{align}
 & \alpha=2.94-\log[(10^{X+2.33})^{-1.08}+(10^{X+2.33})^{2.29}], \\
 & \beta=4.23+1.34X+0.26X^{2}, \\
 & \gamma=-0.06+\log[(10^{X+2.56})^{-0.68}+(10^{X+2.56})].
\end{align}
We refer to this as the DC14 model.
For both CDM models, we adopt the same priors on halo mass and concentration, and use the same MCMC sampling strategy as in the SIDM analysis to enable direct comparison.

We compare the two CDM baselines with our SIDM model using the reduced chi-square, $\chi_{\nu}^2$, and the Bayesian Information Criterion \citep[BIC,][]{Schwarz78},
\begin{equation}
    \text{BIC}=-2\ln\mathcal{L}+k\ln n,
\end{equation}
where $k$ is the number of free parameters, $n$ is the number of data points, and $\mathcal{L}$ is the maximum likelihood. 
Models with smaller $\chi_{\nu}^2$ and BIC values are preferred. 
For SIDM double-solution systems, we adopt the better-fit solution.

As shown in \Fig{BIC}, for about 80\% of the sample, the SIDM model yield slightly better fits than the DC14 CDM model in terms of $\chi_\nu^2$, and both outperform the contracted NFW model.  
The difficulty for the contracted NFW profile lies in producing central density cores, whereas DC14 (through baryonic feedback) and SIDM (through self-scattering) can generate a spectrum of cusps and cores. 
The BIC comparison similarly disfavors contracted NFW relative to SIDM, and the SIDM and DC14 model have similar performance. 
This is because SIDM introduces additional free parameters (the cross-section parameters), while in DC14 the three profile slopes are tied directly to the stellar-to-halo mass ratio $X$, and thus do not increase the model’s formal degrees of freedom. 

\subsection{Dark self-interaction or baryonic feedback?}
\label{sec:SIDMorFeedback}

\begin{figure}
    \centering
    \includegraphics[width=0.72\linewidth]{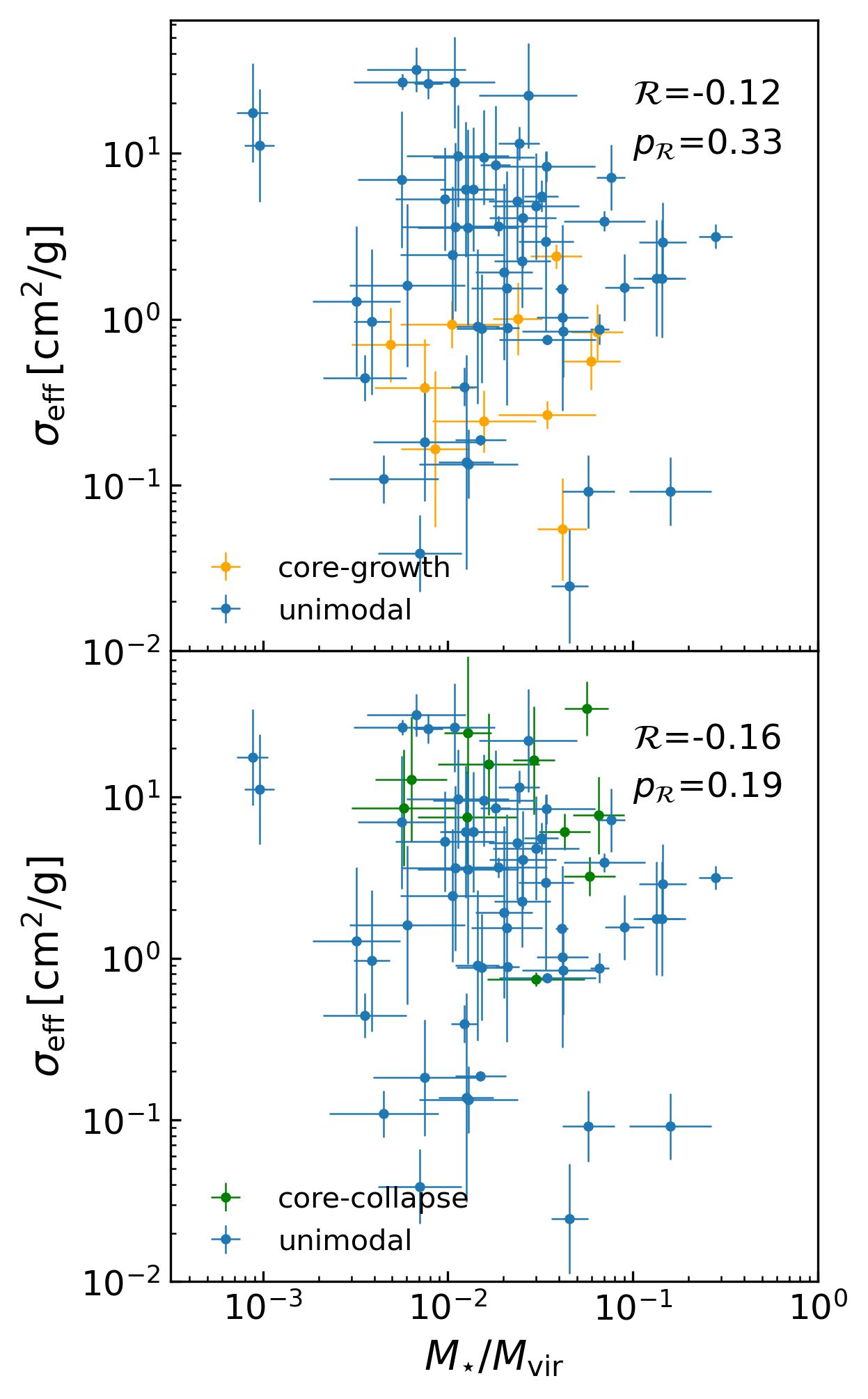}
    \caption{
    Correlation between the SIDM effect, quantified by the effective cross section $\sigmaeff$, and baryonic effects, proxied by the stellar-to-halo mass ratio $\Ms/\Mv$. 
    In the {\it upper} and {\it lower} panels, the orange and green symbols represent the core-growth and core-collapse solutions, respectively, for galaxies with bimodal posteriors.
    Galaxies with unimodal posteriors are shown in blue.  
    The Spearman correlation coefficients $\mathcal{R}$ and their associated $p_\mathcal{R}$ values are reported in each panel. 
    \quad Overall, the correlations are weak to absent. These results suggest that SIDM introduces a distinct source of structural diversity in halos, largely independent of baryonic feedback effects.
    }
    
    \label{fig:SIDMvsFeedback}
\end{figure}

One of the main effects of SIDM is the formation of density cores. However, cores can also be produced by baryonic feedback, particularly through gas outflows driven by supernovae \citep[e.g.,][]{Freundlich20}. 
To rigorously disentangle these contributions, one would ideally employ an SIDM model that self-consistently includes baryonic feedback and allows each mechanism to be switched on or off in controlled experiments. While such an approach is beyond the scope of this work, we can still gain qualitative insight through the following analysis.

We ask whether the SIDM effect is degenerate with baryonic feedback.
To do so, we take the effective cross section, $\sigmaeff$, as a proxy for the strength of SIDM, and the stellar-to-halo mass ratio $\Ms/\Mv$ as a proxy for feedback strength.
The reasoning is straightforward: baryonic core formation is expected to be strongest near $\Ms/\Mv\sim0.01$ \citep[e.g.,][]{DC14}. If $\sigmaeff$ shows a local maximum in this regime, it would suggest that SIDM effects overlap or degenerate with baryonic feedback. 
Conversely, a lack of correlation would point to SIDM capturing distinct physical processes.

As shown in \Fig{SIDMvsFeedback}, we find no evidence for such a local maximum, and the correlation between $\sigmaeff$ and $\Ms/\Mv$ is essentially absent, with the Spearman correlation coefficient $\mathcal{R}$ close to zero and its $p_{\mathcal{R}}$ value large. 
We therefore conclude that SIDM introduces a distinct channel for structural diversity in halos, separate from the effects of baryonic feedback.

\subsection{SIDM impact on scaling relations}
\label{sec:ScalingRelations}

\begin{figure*}
    \centering
    \includegraphics[width=\linewidth]{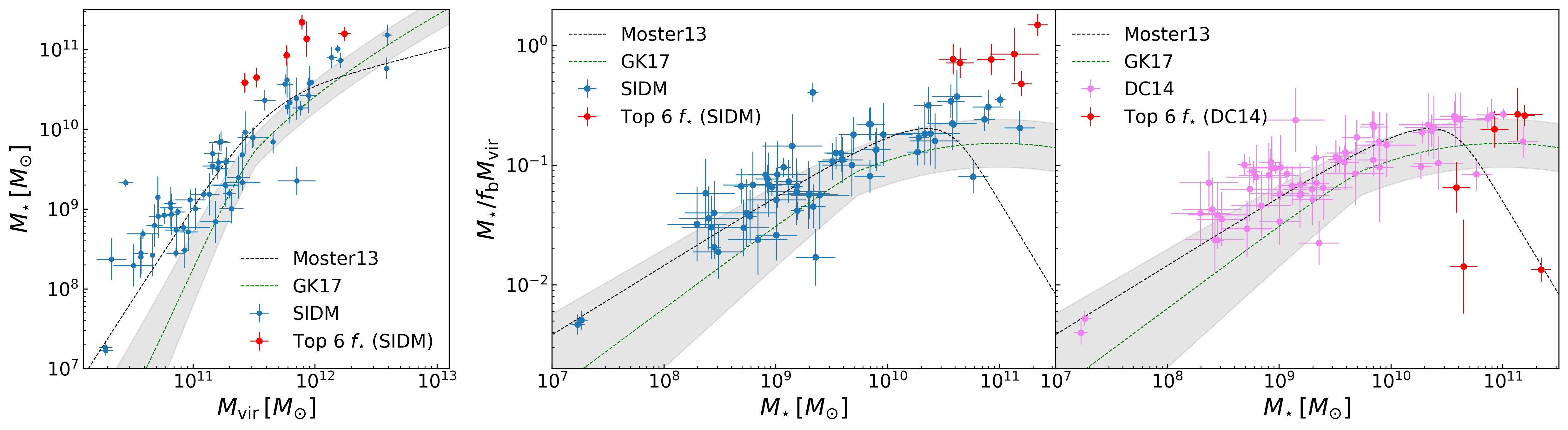}
    \caption{
    Stellar-mass-total-mass relations (SHMRs) shown as $\Ms$ versus $\Mv$ ({\it left}), and as the star-formation-efficiency ratio $f_{\star}\equiv\Ms/(\fb\Mv)$ versus $\Ms$ ({\it middle} and {\it right}). 
    For simplicity, we focus on the core-growth solutions for galaxies with bimodal posteriors, together with  
    unimodal systems. 
    The right panel compares our SIDM results to those obtained under the CDM framework using the \citet{DC14} model.
    \quad Relative to the reference relations from \citet{Moster13} (black line) and \citet{GK17} (green line, with the shaded band marking the 1$\sigma$ scatter), SIDM fits systematically yield lower halo masses and correspondingly higher $f_{\star}$ at the high‑stellar‑mass end, whereas DC14 results remain broadly consistent with the CDM expectations. Red circles in all panels mark the six galaxies with the highest $f_{\star}$ in our SIDM fits.
    We note that a similar discrepancy at high-stellar-mass end has been reported in earlier CDM-based studies of disc galaxies \citep{Posti19}, which suggests that the tension is not resolved by introducing SIDM. Our results thus reaffirm the existence of high-$f_{\star}$ outliers even within an SIDM scenario.
    }
    \label{fig:SHMR}
\end{figure*}

With a statistical sample, we are able to probe the stellar-to-halo mass relation (SHMR) in the SIDM framework and compare it directly with CDM results.
For simplicity, we adopt the core-growth solution for galaxies with bimodal posteriors.  
The left panel of \Fig{SHMR} shows the SHMR for the 68 SPARC galaxies in our analysis.
For reference, we include the abundance-matching relations of \citet{Moster13} and \citet[][adopting the growing-scatter model for field galaxies]{GK17}, both based on $\Lambda$CDM cosmological simulations. 
These two CDM relations differ by nearly 0.5 dex in halo mass at fixed stellar mass, underscoring the substantial uncertainty in the dwarf-galaxy regime of the SHMR even within $\Lambda$CDM.
We identify six galaxies that exhibit the highest total star‑formation efficiency $f_{\star}\equiv\Ms/(\fb\Mv)$ in our SIDM fits, where $\fb\equiv0.188$ is the cosmological baryon fraction. 
In the  left panel of \Fig{SHMR}, these systems (red points) reside in halos that are $\sim 0.5$ dex less massive than predicted by the CDM reference relations at the same stellar mass.

The middle and right panels of \Fig{SHMR} recast the SHMR in the form of $f_{\star}$ as a function of stellar mass, and compare the SIDM and DC14 results side by side. 
Both fits are broadly consistent with the \citet{Moster13} for bright dwarfs ($\Ms \approx 10^7-10^9 \Msun$). 
However, the same six high-$f_{\star}$ galaxies show values close to unity in the SIDM case, which is a significant deviation from both the \citet{Moster13} and \citet{GK17} relations. 
This discrepancy is not seen in the DC14 fits, where galaxies at the high‑mass end remain consistent with the CDM SHMR.

We note that a similar departure from the SHMR for disc galaxies has been reported in earlier CDM‑based studies \citep{Posti19}. Notably, the two galaxies with $f_{\star}>1$ identified in \citet{Posti19}, NGC5371 and NGC0801, correspond to the two systems with the highest $f_{\star}$ in our present analysis.
Hence the deviation does not uniquely distinguish between dark‑matter scenarios, but indicates that the tension persists even when a new dark‑matter species is assumed. We suspect that an overestimated mass-to-light ratio leads to the departure observed in both \citet{Posti19} and our work. To examine the origin of the difference between SIDM and DC14 fits, we also mark the same six galaxies with red circles in the right panel of \Fig{SHMR}. 
Among them, three have unreliable mass constraints $\Vvir/\Vmax \gtrsim1$ and also show the lowest $f_{\star}$ values within the group. The remaining three galaxies have virial masses roughly 0.5 dex larger in the DC14 fit than in the SIDM fit.

\begin{figure}
    \centering
    \includegraphics[width=0.8\linewidth]{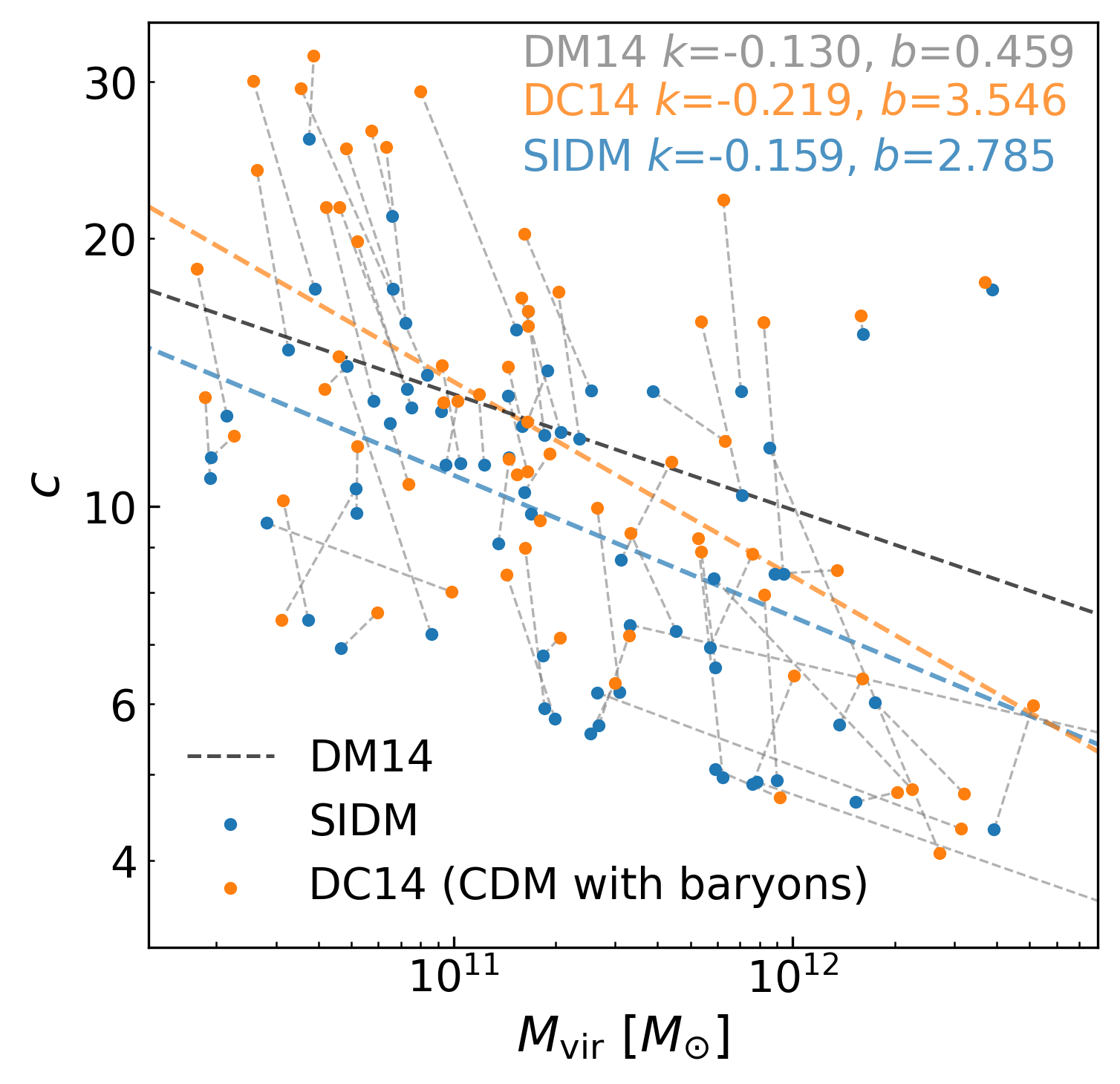}
    \caption{
    Halo concentration-mass relation derived from our SIDM mass models and the \citet{DC14} CDM models. 
    For galaxies with bimodal SIDM posteriors, we adopt the core-growth solution.
    CDM and SIDM results for the same galaxy are connected by gray dashed lines.
    The black dashed line marks the \citet{DM14} concentration-mass relation, and the blue and orange dashed lines respectively represent the best-fit concentration–mass relations derived from the SIDM and DC14 fits, with their slopes and intercepts provided in the panel. We find that SIDM yields systematically lower concentrations than the DM14 reference relation, albeit with a similar slope. In contrast, the DC14 fits result in both a steeper slope and a higher normalization of the relation.
    }
    \label{fig:concentration}
\end{figure}

In addition, we examine the halo concentration - halo mass relation, again, adopting the core-growth solution in bimodal cases for SIDM fits.
The results are shown in \Fig{concentration}. 
The reference line indicates the widely used empirical concentration-mass relation of \citet{DM14} (DM14). 
Since this relation is included as a prior for both the SIDM and DC14 fits, it is not surprising that the resulting $c$-$\Mv$ relations broadly follow the input trend.
However, a notable difference emerges: the SIDM mass models produce a systematically lower concentration than the DM14. 
To quantify this offset, we assume $\log c$ obeys Gaussian distribution, with a constant standard deviation $\sigma$ and the median described by a linear relation, $\log c=k\log\Mv+b$, 
where $\sigma$, $k$, and $b$ are free parameters. 
We fit this simple relation to the results of both CDM and SIDM models. 
As clearly visible in \Fig{concentration}, the median SIDM relation is below DM14, while the slope is similar. The fit to DC14 results shows a steeper slope and a higher normalization. 
Both SIDM and DC14 results show slightly larger scatters, $\sigma_{\mathrm{SIDM}}=0.159$ and $\sigma_{\mathrm{DC14}}=0.176$ respectively, compared to DM14 ($\sigma_{\mathrm{DM14}}=0.11$).

\section{Conclusion}
\label{sec:conclusion}

In this study, we have improved the semi-analytical isothermal Jeans model for computing self-interacting dark matter (SIDM) halo profiles, and applied it to constrain the self-scattering cross section using rotation curves from the SPARC survey. 
Our advances include incorporating velocity-dependent cross sections through an effective cross section formalism \citep{Yangyu22}, improving numerical stability over the previous implementation \citep{Jiang23}, and adopting an empirical treatment of gravothermal core collapse \citep{YangJiang24}. 
These improvements allow us to model a larger and more diverse galaxy sample, including systems with cuspy or baryon-dominated centers that were often excluded in earlier studies.

Analyzing 68 SPARC galaxies, we find that about 1/6 exhibit bimodal posteriors in the parameter space spanned by halo mass, concentration, and effective cross section (or rather, the product of effective cross section and halo age). 
This bimodality arises because core-growth and core-collapse halos of similar mass and concentration can yield similar density profiles. 
We interpret this as tentative evidence for the presence of core-collapsing halos in the SIDM framework. Among unimodal systems, about 50\% are consistent with being at the maximal-core stage, while the rest are best fit in the core-expanding phases.

From these models we derive constraints on the velocity-dependent cross-section parameters, $\sigma_0$ and $\omega$.
Two scenarios were considered depending on how bimodal systems are treated.
The effective cross section $\sigmaeff$ anti-correlates strongly with the maximum circular velocity $V_{\text{max}}$, 
yielding a well-defined L-shaped degeneracy in the $\sigma_0$–$\omega$ plane, and the shape is not sensitive to the choice of solutions for bimoldal cases. 
Both nearly constant cross sections ($\sigma_0\sim 2 \cm^2/\g$, $\omega \gtrsim 500\kms$) and strongly velocity-dependent ones ($\sigma_0\sim 100\cm^2/\g$, $\omega\sim 60\kms$) are comparably viable.
Fitting the ($\Vmax$,$\sigmaeff$) data gives representative best-fit parameters of $(\sigma_0,\omega)$ = $(4.9\cm^2/\g, 250\kms)$ or $(6\cm^2/\g, 220\kms)$, depending on the assumed evolutionary branch for the bimodal galaxies.

We find that the parameter space favored by our model is broadly consistent with previous constraints from nearby galaxies, while also encompassing values that independently explain the supermassive black hole mass functions inferred for the Little Red Dots at high redshift \citep{Jiang25}. 
Taken together, these results highlight a region of SIDM parameter space that simultaneously satisfies orthogonal constraints from local galaxy dynamics and black-hole demographics in the early Universe.

We also compare our SIDM model against two cold dark matter (CDM) models that incorporate baryonic effects: a contracted NFW profile and the empirical \citet[][DC14]{DC14} model. 
SIDM and DC14 outperform the simple contraction model, and have similar performance in BIC.
Given that CDM simulations still disagree on the strength of core formation driven by baryonic feedback, the fact that SIDM naturally generates both cores and cusps without invoking fine-tuned baryonic physics remains a key advantage. 
Moreover, the effective cross sections from our fits show marginal correlations with the stellar fraction -- a proxy for feedback strength, suggesting that SIDM accounts for structural diversity in a way distinct from baryonic feedback.  

Finally, we explore implications for galaxy–halo scaling relations. 
The stellar-to-halo mass relation (SHMR) from SIDM fits shows that the total star-formation efficiency  $\Ms/(\fb\Mv)$ of some galaxies exceeds the prediction of CDM SHMR and, in a couple of cases, close to unity. 
This is not unique to the SIDM mass models: previous CDM-based mass models by \citet{Posti19} reveal similar cases. 
We conclude that SIDM does not significantly vary the SHMR. 
In addition, we find that the SIDM concentration–mass relation systematically falls below the CDM relation \citep{DM14}

In summary, our analysis demonstrates that SIDM can successfully reproduce the diversity of galaxy rotation curves, but also highlights key degeneracies and the broad scatter in inferred cross sections. 
The joint constraints consistently favor velocity-dependent interactions, with the characteristic L-shaped degeneracy in the $(\sigma_0, \omega)$ plane. 
These results underscore both the potential and the limitations of current SIDM modeling: the framework captures essential features of galactic kinematics, yet no single cross-section model fits the entire galaxy sample. 
Breaking these degeneracies using nearby kinematics will require improved theoretical treatments and new kinematic data, especially for ultra-faint dwarfs at $V_{\rm max} \lesssim 30 \kms$. 


\section*{Acknowledgements}
We thank Haibo Yu, Daneng Yang, and Luis Ho for insightful comments and suggestions. 
This work is supported by the China Manned Space Program
with Grant No. CMS-CSST-2025-A03.
FJ acknowledges support from the National Natural Science Foundation of China (NSFC, Grant No. 12473007). 
ZJ acknowledges support from the Beijing Natural Science Foundation (Grant No. QY23018). 
RL acknowledges the support of NSFC (Grant No. 11988101) and CAS Project for Young Scientists in Basic Research (No. YSBR-062).  
LZ acknowledges the support from CAS Project for Young Scientists in Basic Research, Grant No. YSBR-062 and 
National Key R\&D Program of China No. 2025YFF0511000.

\section*{Data Availability}

The improved isothermal Jeans model developed in this work is publicly available at: 
\url{https://github.com/ZixiangJia/SIDM_Jeans_model}, 
and the version in development is available upon request to the corresponding author. 
The data used in the analysis is publicly available at the SPARC survey website: 
\url{https://astroweb.cwru.edu/SPARC/}. 
The fitting results can be found at \url{https://github.com/ZixiangJia/Jia26_fitting_results}.



\bibliographystyle{mnras}
\bibliography{bibliography} 




\appendix

\section{Baryonic Effect on the Velocity Distribution}
\label{apdx: baryonic effects on the velocity distribtion}

Baryons deepen the gravitational potential of dark matter halos and modify the inner velocity dispersion \citep{ZhongYang23}. 
As a result, $\Vmax$ of the initial NFW profile may no longer accurately trace the inner velocity dispersion of the halo. 
In this appendix, we demonstrate that the effect of baryons is secondary for the galaxy samples examined in our work.

\begin{figure}
    \centering
    \includegraphics[width=0.8\linewidth]{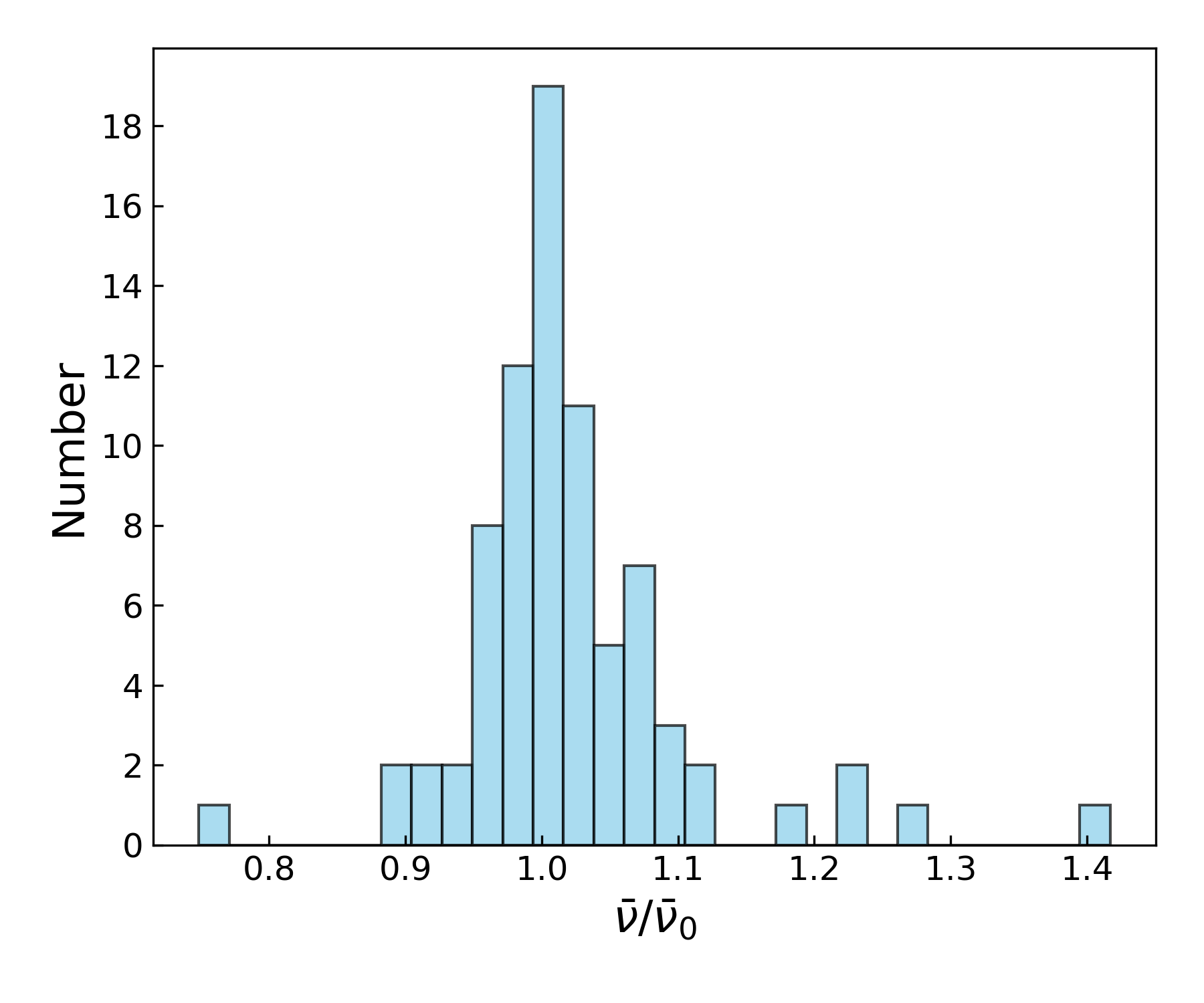}
    \caption{Distribution of the velocity dispersion ratio for the galaxy samples in this work. 
    $\bar{\nu}$ and $\bar{\nu}_0$ are the averaged velocity dispersions within  $r_1$ after and before adiabatic contraction, respectively. 
    Most systems show a fractional change within 10\%, indicating that baryons do not alter the inner velocity dispersion significantly.}
    \label{fig:sigma_ratio_distribution}
\end{figure}

Specifically, we compute the ratio of the velocity dispersion averaged within $r_1$
after adiabatic contraction, $\bar{\nu}$, to that before adiabatic contraction, $\bar{\nu}_0$. 
The velocity dispersion is averaged as follows
\begin{equation}
    \bar{\nu}^2 = \frac{1}{M(r_1)}\int_0^{M(r_1)}\nu^2(r)\mathrm{d}M(r),
\end{equation}
where $\nu(r)$ and $M(r)$ is the velocity dispersion and enclosed mass at radius $r$, respectively. 
\Fig{sigma_ratio_distribution} shows the distribution of this ratio for our galaxy samples. 
For the majority of the samples, the fractional change in the velocity dispersion is less than 10\%. 
This suggests that baryons do not substantially alter the inner velocity dispersion, and  $\Vmax/\sqrt{3}$
based on the initial NFW profile remains a valid estimate of the characteristic velocity dispersion.
Consequently, we ignore this effect throughout our work.

\section{Improvement on the isothermal Jeans model}
\label{app:model}
\begin{figure*}
    \centering
    \includegraphics[width=1\linewidth]{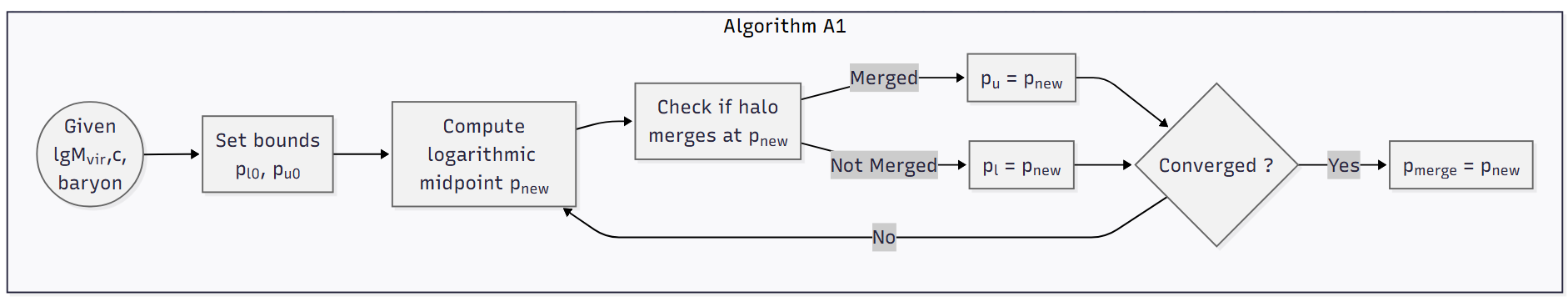}
    \caption{Flowcharts of algorithm A1 (details in \app{find tmerge}), which implements a bisection method for determining the halo merge point $p_{\text{merge}}$. Starting from initial bounds $p_{\text{l0}}$ (pre-merger) and $p_{\text{u0}}$ (post-merger), the algorithm iteratively narrows the interval containing $p_{\text{merge}}$ by testing whether the isothermal Jeans solutions have merged at the logarithmic midpoint $p_{\text{new}}$. This step is essential for subsequently approximating the density profile of a core-collapsed halo via the high-density-solution-mirroring technique.}
    \label{fig:algorithm}
\end{figure*}

In this appendix, we detail the improvements on the Jeans model. To extend the Jeans model into the core‑collapse regime, following  \citet{YangJiang24}, we approximate the density profile of a core‑collapsed halo with a given product $p$ using the high‑density solution evaluated at $p'=2p_{\text{merge}}-p$. To implement this, we develop an efficient algorithm that accurately determines the merge point $p_{\text{merge}}=\sigmaeff\tmerge$. We further refine the stitching procedure by introducing a criterion to exclude non‑physical outcomes and performing multiple searches across partitioned intervals in parameter space. This enhancement allows robust identification of both the low‑density and high‑density solution branches. Together, these improvements support fast and reliable computation  of the Jeans model across all stages of halo evolution.

\subsection{Calculation of the merge point of the halo}
\label{app:find tmerge}
We employ a bisection method to precisely determine the merge point $\pmerge$ of a halo.  This approach is based on a characteristic feature of the Jeans model: assuming \Eq{r1}, when $p > p_{\text{merge}}$, the model yields only physically invalid solutions, whereas for $p < p_{\text{merge}}$, at least one valid solution can be obtained through sufficient searching. To quantify this behavior, we propose that the objective quantity
\begin{equation}
\delta^2 = \left[ \frac{\rho_{\text{iso}}(r_1) - \rho_{\text{cdm}}(r_1)}{\rho_{\text{cdm}}(r_1)} \right]^2 + \left[ \frac{M_{\text{iso}}(r_1) - M_{\text{cdm}}(r_1)}{M_{\text{cdm}}(r_1)} \right]^2,
\end{equation}
which quantifies the deviation from the boundary conditions required by the Poisson equation, serves as an effective indicator of invalid solutions. Empirically, we find that $\delta^2 < 10^{-5}$ provides a reliable threshold for selecting physically meaningful solutions. Leveraging this clear transition and the aforementioned characteristic behavior of the model, we construct a bisection algorithm to precisely locate $p_{\text{merge}}$, with the flowchart of the algorithm shown in \Fig{algorithm}.

The algorithm proceeds as follows. We first select an initial lower bound $p_{\text{l0}}$ (in the pre-merger regime) and an upper bound $p_{\text{u0}}$ (in the post-merger regime). The logarithmic midpoint  is then calculated as $\log p_{\text{new}} = (\log p_{\text{l0}} + \log p_{\text{u0}})/2$. To exclude the cases  where a single search fails to find a valid solution when $p<\pmerge$, we perform four independent searches with different initial guesses and  evaluate their $\delta^2$. A halo is considered to have merged $(p>\pmerge)$ only if all four searches return $\delta^2 > 10^{-5}$. After assessing the halo state at $p_{\text{new}}$, if the halo has merged, we update the upper bound to $p_{\text{u1}} = p_{\text{new}}$ while retaining the lower bound $p_{\text{l1}} = p_{\text{l0}}$. If it has not merged, the lower bound is updated to $p_{\text{l1}} = p_{\text{new}}$ and the upper bound remains $p_{\text{u1}} = p_{\text{u0}}$. This process iterates until the bounds satisfy a specified convergence criterion, at which point the final $p_{\text{new}}$ is returned as the merge point. The entire bisection scheme enables the precise determination of a halo's merge point within seconds.

\subsection{Robustly identifying the solution}
\label{app:identify solutions}
The stitching process occasionally converges to saddle points or parameter space boundaries when $p<\pmerge$, leading to an invalid solution. To robustly identify physically valid solutions across all evolutionary stages of the halo, we partition the search intervals and perform multiple searches from different initial positions. This approach substantially improves the robustness of the solution identification and ensures reliable outputs.

\begin{figure}
    \centering
    \includegraphics[width=0.8\linewidth]{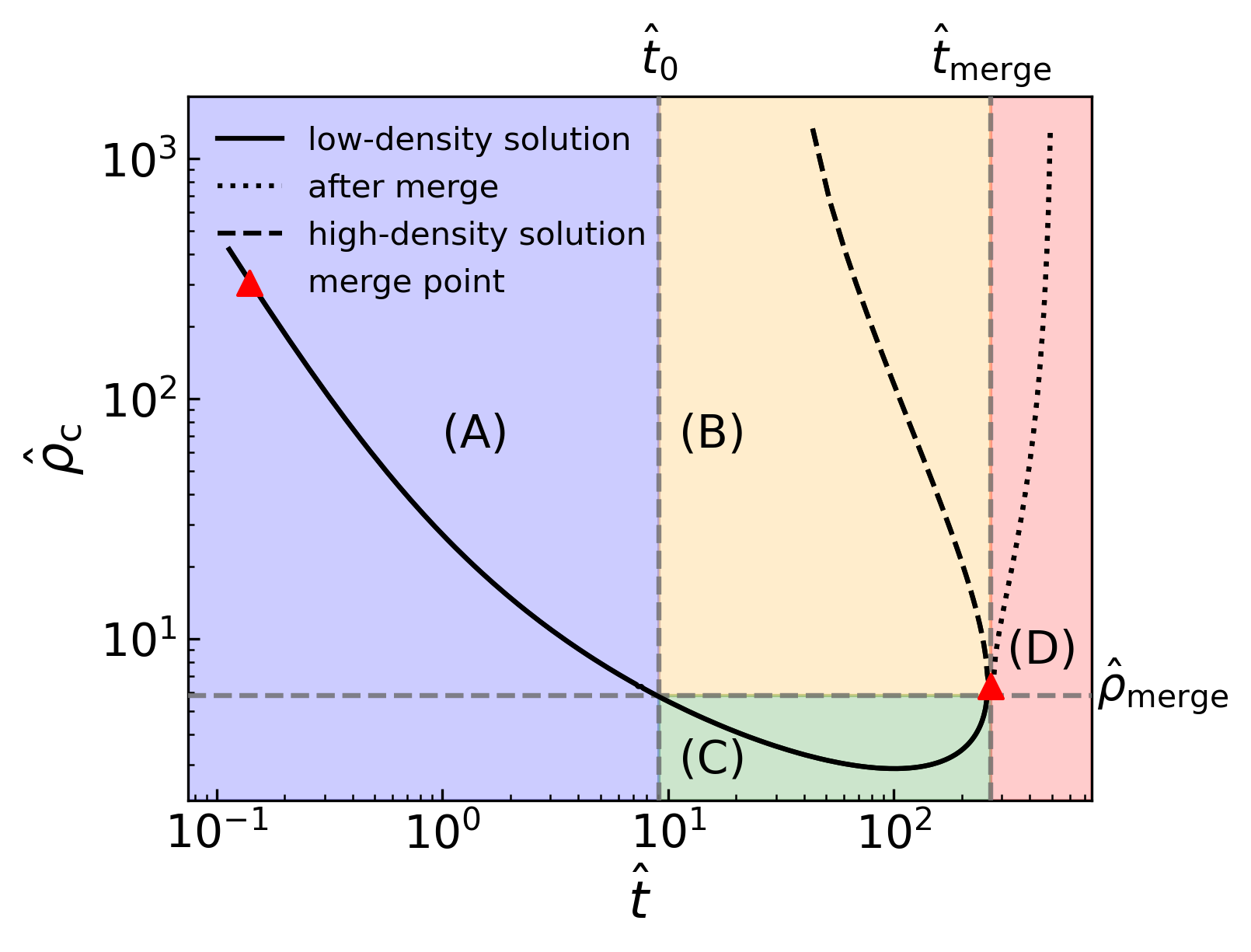}
    \caption{Gravothermal evolutionary stages of an SIDM halo, divided based on the normalized central density $\hat{\rho}$ as a function of the normalized age $\hat{t}$. Two dashed vertical lines denote the transition boundaries $\hat{t}_0$ and $\hat{t}_{\rm merge}$, separating the evolution into the three distinct regimes employed in the Jeans model. Here, $\hat{t}_0$ corresponds to the point where the normalized central density $\hat{\rho}_{\mathrm{c}}$ equals its value at the merge point  $\hat{\rho}_{\rm merge}\equiv\hat{\rho}_{\mathrm{c}}(\hat{t}=\hat{t}_{\mathrm{merge}})$, with  $\hat{t}_{\mathrm{merge}}$ representing the normalized age at $\pmerge$.}
    \label{fig:different stages}
\end{figure}

For halos at different evolutionary stages, we employ distinct search strategies. We define the normalized halo age as $\hat{t}\equiv8\sqrt{G\rhos}\rhos \rs \sigmaeff \tage\propto p$, and let $\hat{t}_{0}$ denote the point at which the normalized central density $\hat{\rho}_{\mathrm{c}}\equiv\rho_{\mathrm{c}}/\rho_{\mathrm{s}}$ equals that at the merge point $\hat{\rho}_{\rm merge}\equiv\hat{\rho}_{\mathrm{c}}(\hat{t}=\hat{t}_{\mathrm{merge}})$, where $\hat{t}_{\mathrm{merge}}$ is the normalized age corresponding to  $\pmerge$. The evolution of an SIDM halo is divided into three stages based on $\hat{t}_0$ and $\hat{t}_{\mathrm{merge}}$, as illustrated in \Fig{different stages}.

When $\hat{t}\leq\hat{t}_0$  (region A in \Fig{different stages}), the corresponding mirrored solution with $\hat{t}' = 2\hat{t}_{\mathrm{merge}} - \hat{t}$ would lie deep in the collapsed regime, beyond the predictive range of the Jeans model. Hence we assume that a high‑density branch does not exist in this stage. In this case,  we require the lower limit of the search interval of $\rhodmc$ to be $\rho_{\mathrm{merge}}$, the central density at $\hattmerge$. 

When $\hat{t}_0\leq\hat{t}<\hattmerge$ (region C), the model is expected to yield the low‑density solution. Here we impose $\rho_{\text{merge}}$ as the upper bound of the $\rhodmc$ search interval.

When $\hat{t}\geq\hattmerge$ (region D), we approximate the density profile using the high‑density solution evaluated at the mirrored age $\hat{t}'=2\hattmerge-\hat{t}$, which lies in  region B. For this regime, the lower bound of the $\rho_{\text{dm0}}$ search is again set to $\rho_{\text{merge}}$. 

For a given halo, we first check whether its product $p$ exceeds $\pmerge$. If so, the solution of the Jeans model lies in region D; in this case, we approximate it using the high-density solution in region B. Otherwise, we examine whether the solution resides in region C by constraining the upper bound of the search interval for $\rhodmc$ to $\rho_{\mathrm{merge}}$. If all search attempts return physically invalid outcomes, we conclude that the solution belongs to region A and execute the corresponding search strategy accordingly.

In all regimes, search for $(\rhodmc,v_0)$ is performed with up to 20 independent attempts from different initial guesses. Solutions that yield $\delta^2 > 10^{-5}$ are rejected as physically invalid. This multi‑search strategy, combined with the stage‑dependent bounds, ensures robust identification of the appropriate solution branch across the entire evolution of an SIDM halo.

\section{Mass-to-light ratio}
\label{apdx: mass-to-light ratio}

\begin{figure}
    \centering
    \includegraphics[width=0.75\linewidth]{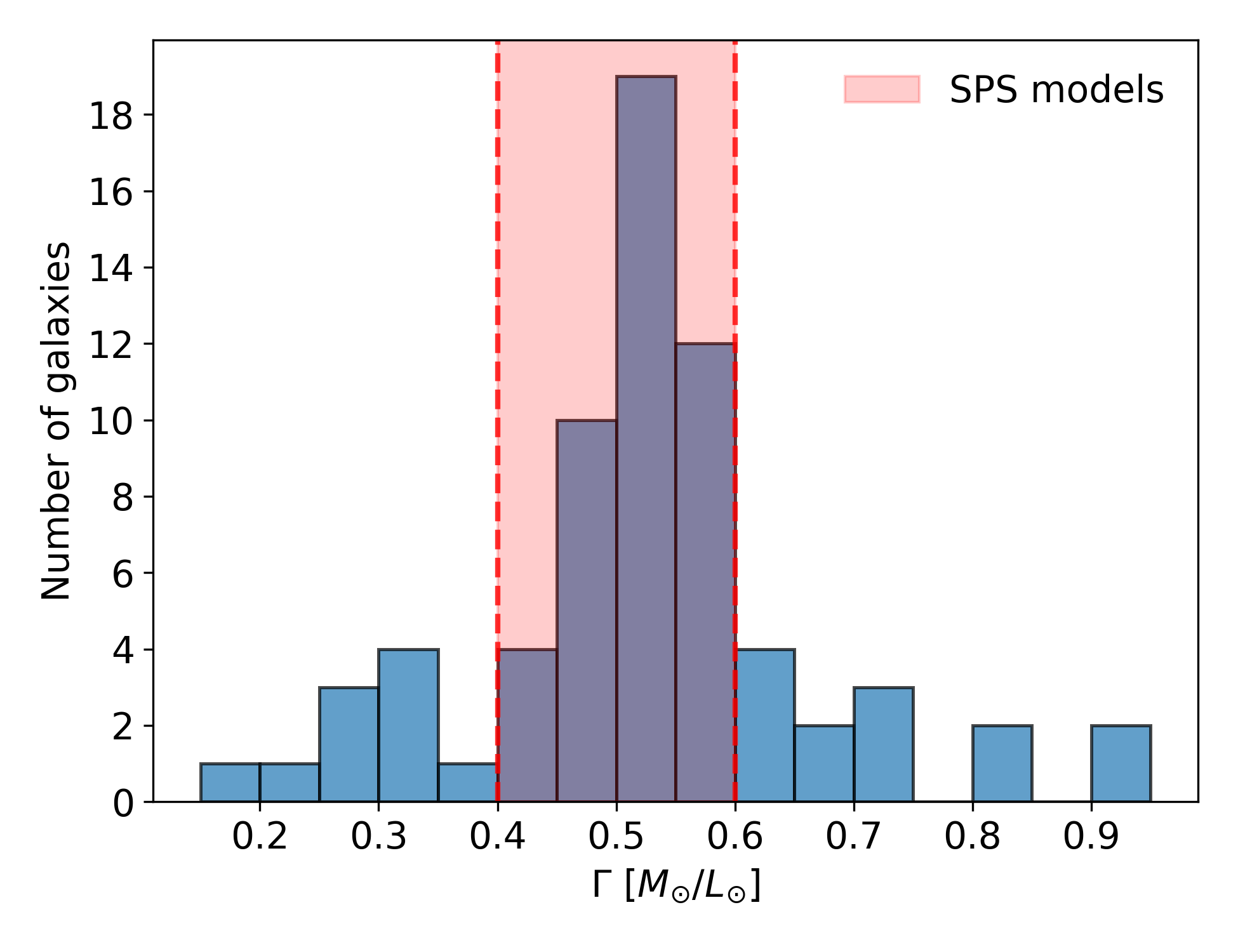}
    \caption{
    Distribution of mass-to-light ratio $\Gamma$ at 3.6$\,\mu$m for the 68 galaxies in our study. 
    The vertical shaded band marks the range $\Gamma = 0.4-0.6\,\Msun/L_{\odot}$, as predicted by stellar population synthesis (SPS) models, depending on the model and initial mass function \citep{Lelli16}. 
    The histogram shows that the majority of galaxies lie within this theoretical range, confirming that our adopted mass-to-light ratio values are generally consistent with SPS expectations.
    }
    \label{fig:mass-to-light ratio}
\end{figure}

In this appendix, we demonstrate that the mass-to-light ratio derived in this study is generally  consistent with the predictions of stellar population synthesis (SPS) models. 
According to \citet{Lelli16}, SPS models yield an average mass-to-light ratio at $3.6\,\mu$m in the range of 0.4-0.6$\,\Msun/L_{\odot}$, depending on the specific model and initial mass function (IMF). 
In \Fig{mass-to-light ratio}, we present the distribution of the mass-to-light ratios for the 68 galaxies in our study. 
As shown, the majority of galaxies fall within the range predicted by SPS models, which shows general agreement with theoretical expectations.

\bsp	
\label{lastpage}
\end{document}